\documentclass{raa_twocolumn}      
\usepackage{graphicx,times}
\usepackage{natbib}
\usepackage{amssymb,amsmath}
\usepackage{upgreek}
\bibpunct{(}{)}{;}{a}{}{,}
\usepackage[pagebackref=true]{hyperref}
\usepackage{supertabular}
\usepackage{threeparttable}
\usepackage{caption}

\begin{document}

\title{Time-resolved Spectral Properties of Fermi-GBM Bright Long Gamma-Ray Bursts}

\volnopage{ {\bf 20XX} Vol.\ {\bf X} No. {\bf XX}, 000--000}

\setcounter{page}{1}

\author{Wan-Kai Wang\inst{1,2}, Wei Xie\inst{1,2,}$^{\ast}$, Zhi-Fu Gao\inst{3}, Shuo Xiao\inst{1,2}, Ai-Jun Dong\inst{1,2}, Bin Zhang\inst{1,2}, Qi-Jun Zhi\inst{1,2}
   }

\institute{Department of Astronomy, School of Physics and Electronic Science, Guizhou Normal University, {\it Guiyang, 550001, China}; {\it xieweispring@gznu.edu.cn}\\
        \and
            Guizhou Provincial Key Laboratory of Radio Astronomy Data Processing, Guizhou Normal University, {\it Guiyang, 550001, China}\\
        \and 
            Xinjiang Astronomical Observatory, Chinese Academy of Sciences, {\it Urumqi, Xinjiang, 830011, China}\\
\vs \no
   {\small Received 20XX Month Day; accepted 20XX Month Day}
}

\abstract{The prompt emission mechanism of gamma-ray bursts (GRBs) is still unclear, and the time-resolved spectral analysis of GRBs is a powerful tool for studying their underlying physical processes. We performed a detailed time-resolved spectral analysis of 78 bright long GRB samples detected by Fermi/Gamma-ray Burst Monitor (GBM). A total of 1490 spectra were obtained and their properties were studied using a typical Band-shape model. Firstly, the parameter distribution of the time-resolved spectrum given as follows: the low-energy spectral index $\alpha \sim -0.72$, high-energy spectral index $\beta \sim -2.42$, the peak energy $E_{\rm p} \sim 221.69 \,\rm{keV}$, and the energy flux $F \sim 7.49\times 10^{-6} \rm{\, erg\,cm^{-2}\,s^{-1}}$. More than 80\% of the bursts exhibit the hardest low-energy spectral index $\alpha_{\rm max}$ exceeding the synchrotron limit (-2/3). Secondly, the evolution patterns of $\alpha$ and $E_{\rm p}$ were statistically analyzed. The results show that for multi-pulse GRBs the intensity-tracking pattern is more common than the hard-to-soft pattern in the evolution of both $E_{\rm p}$ and $\alpha$. The hard-to-soft pattern is generally shown in single-pulse GRBs or in the initial pulse of multi-pulse GRBs. Finally, we found a significant positive correlation between $F$ and $E_{\rm p}$, with half of the samples exhibiting a positive correlation between $F$ and $\alpha$. We discussed the spectral evolution of different radiation models. The diversity of spectral evolution patterns indicates that there may be more than one radiation mechanism occurring in the gamma-ray burst radiation process, including photospheric radiation and synchrotron radiation. However, it may also involve only one radiation mechanism, but more complicated physical details need to be considered.
}

   \authorrunning{W.-K. Wang et al. }            
   \titlerunning{Time-resolved Spectral Properties of Fermi-GBM Bright Long Gamma-Ray Bursts}  
   \maketitle

%
\section{Introduction}           
\label{sect:intro}

Gamma-ray bursts (GRBs) are the most intense explosions in the universe, which have been extensively studied since their discovery in the 1960s. However, there are still several unidentified issues during the prompt emission phase (\citealt{Zhang+etal+2011, Kumar+etal+2015}). 
For example, the composition of the jet (baryonic matter or Poynting flux), the energy dissipation and particle acceleration mechanisms (internal shocks or magnetic reconnection, see \citealt{Zhangbin+etal+2011}), the radiation mechanisms (synchrotron radiation (\citealt{Meszaros+etal+1994, Ravasio+etal+2018}), inverse Compton scattering (\citealt{Kumar+etal+2008}), and photosphere radiation (\citealt{Meng+etal+2022}). The lightcurves of GRB exhibit irregularity and diversity, while the spectra tend to have a few simpler shapes compared to the lightcurves, which can be well described by several experimental functions. The most typical of them are the Band function (Band, see \citealt{Band+etal+1993}), cutoff power law (CPL), single power law function (PL), and smoothly broken power law function (SBPL). The above spectral models are usually considered non-thermal spectra. In some cases, the spectra of GRBs are better fitted by superimposing multiple spectral functions rather than by a single spectra function, such as an additional power-law component  (\citealt{Abdo+etal+2009, Ackermann+etal+2010}) or a thermal component (\citealt{Ghirlanda+etal+2003, Ryde+etal+2004, Ryde+etal+2005, Ryde+etal+2009, Guiriec+etal+2011, Axelsson+etal+2012, Li+etal+2019+apjs.b,  Zhao+etal+2022}) on the basis of a Band spectrum. \cite{Ryde+etal+2005} fitted a Planck function superposition power law for some GRBs and found that the thermal component dominates. \cite{Abdo+etal+2009} fitted the spectrum of GRB 090902B with a band function plus power law. \cite{Ryde+etal+2010} claimed that such a burst can also be fitted with a multicolor blackbody function model superimposed on a power law function. The thermal component is located on the left shoulder of the peak energy (below the $E_{\rm p}$) and the power-law component can be extended to the high and sub-low energy bands.\par

The possible radiation mechanisms of the prompt emission phase of GRBs may consist of synchrotron radiation of relativistic electrons, quasi-thermal radiation from the photosphere, as well as the Comptonization of thermal photons (\citealt{Pe'er+etal+2006, Zhang+etal+2011}). Observationally, most of the GRB spectra behave as non-thermal spectra, so the synchrotron mechanism is the preferred model to explain the majority of the prompt emission (\citealt{Baring+etal+2004, Lloyd+etal+2000, Burgess+etal+2014, Zhang+etal+2016}). The spectral shape can be described by the Band function and the CPL function. In general, the low-energy spectral slope is an indicator to determine whether the observed radiation can be correlated with any of the proposed theoretical spectral models. The typical low-energy spectral index $\alpha$ allowed by the internal shock synchrotron model ranges from $-2/3$ (slow-cooling) to $-3/2$ (fast-cooling). Indeed, such a value is much softer than many observations, living the fast-cooling problem of synchrotron radiation (\citealt{Preece+etal+2000}). Although the synchrotron emission in decaying magnetic fields could alleviate the fast-cooling problem (\citealt{Uhm+etal+2014}), the predicted $\alpha$ is still smaller than -2/3. The limit value -2/3 is usually thought as the ``death line" of synchrotron radiation. Spectra with $\alpha$ larger than -2/3 are often considered as emission from the photospheres. However, this criterion can hardly be strict. More than one complex emission model produce a wider range of $\alpha$ values. Models such as the subphotospheric dissipation (\citealt{Rees+etal+2005}) and geometrical broadening in photospheric emission (\citealt{Pe'er+etal+2008}) could efficiently produce a broader spectra. Not only that, the GRBs with $-2/3 \leq \alpha \leq 1$ can be adequately explained with the time-dependent cooling of synchrotron electrons model as well (\citealt{Burgess+etal+2020}). Moreover the “fast cooling problem” can also be reconciled in the Internal-collision-induced Magnetic Reconnection and Turbulence (ICMART) model due to the turbulence heating (\citealt{Zhangbin+etal+2011, Shao+etal+2022}). Therefore, it is difficult to identify the emission model only by $\alpha$.\par

The Band function is characterized by three parameters: the low-energy index $\alpha$, the high-energy index $\beta$, and the peak energy ($E_{\rm p}$). In previous works based on large samples, the typical values of the time-integrated spectrum are $\alpha \sim -1.0$, $\beta \sim -2.0$, and  $E_{\rm p} \sim 300 $ keV (\citealt{Preece+etal+2000, Zhang+etal+2011, Nava+etal+2011, Goldstein+etal+2012}. While for the time-resolved spectrum, the low-energy index is much harder ($\alpha \sim -0.8$, see   
\citealt{Kaneko+etal+2006, Yu+etal+2016, Yu+etal+2019, Li+etal+2021}). For bright bursts, the time-resolved spectral analysis provides more clues about the GRB prompt emission. It has been found that the characteristic parameters of the spectrum are not fixed, but evolve over time (\citealt{Golenetskii+etal+1983, Crider+etal+1997, Kaneko+etal+2006, Peng+etal+2009}), making the spectral evolution pattern become another important indicator for studying the radiation mechanism of GRBs. The evolution characteristics of the time-resolved spectra $E_{\rm p}$ and $\alpha$ have been extensively studied in the early days. During the Compton Gamma Ray Observatory ($CGRO$) observation period, $E_{\rm p}$ showed several different evolution patterns: (i) ``hard-to-soft" (\citealt{Norris+etal+1986}); (ii) ``intensity-tracking" (\citealt{Golenetskii+etal+1983}); (iii) ``soft-to-hard" or chaotic evolution. With the launch of the Fermi Gamma-ray Space Telescope ($Fermi$) in 2008, the quality of the spectral data has been improved, identifying that the first two patterns dominant (\citealt{Lu+etal+2012}). The physical origin of these evolved patterns has not been resolved yet. For single-pulse GRBs, the ``hard-to-soft" pattern accounts for about two-thirds and the ``intensity-tracking" pattern for about one-third (\citealt{Lu+etal+2012, Yu+etal+2016, Yu+etal+2019}). The evolution pattern of multi-pulse GRBs is more complex, showing hard-to-soft followed by intensity-tracking (first pulse ``hard-to-soft" followed by ``tracking") and tracking (including the first pulse) is more common (\citealt{Lu+etal+2012}). Recently, \cite{Li+etal+2021} found that $E_{\rm p}$ showed a tracking pattern in 60\% of their multipulse GRB samples. Analysis of the Fermi-LLE GRB spectrum revealed that the $E_{\rm p}$ ``tracking" behavior accounted for 75\% of the analyzed samples (\citealt{Duan+etal+2020}). As for the $\alpha$ evolution, \cite{Crider+etal+1997} firstly pointed out that it evolves over time, instead of staying the same. The $\alpha$ evolution is more complex and shows multiple evolution patterns. Therefore, there are relatively few studies and physical explanations of the $\alpha$ evolution. In one study, the $E_{\rm p}$ and the $\alpha$ were found to show a ``double tracking" pattern (\citealt{Li+etal+2019+apj}). \par

In order to further investigate the time-resolved spectral properties of bright long GRBs in the prompt emission phase, the time-resolved spectra of 78 bright long GRBs detected by Fermi/Gamma-ray Burst Monitor (GBM) are analyzed in this work. The Band function was used to fit each burst. In addition, their parametric distributions are given by detailed time-resolved spectral analysis. The evolution patterns of $\alpha$ and $E_{\rm p}$ are also statistically analyzed. The correlations of parameters, such as $\alpha-E_{\rm p}$, $F-E_{\rm p}$, and $F-\alpha$, are obtained. 


\section{Sample selection and spectral analysis methods}
\label{sect:Obs}
\subsection{Sample Selection}
Our samples come from the Fermi Gamma-ray Space Telescope, which was launched in June 2008 and carries two instruments. One is the Gamma-ray Burst Monitor (GBM, see \citealt{Meegan+etal+2009}), which observes the energy range of 8\,keV to 40\,MeV, and its main task is to search for GRBs in the universe. The other is the Large Area Telescope (LAT, see \citealt{Atwood+etal+2009}), which has an energy range of 30\,MeV to 300\,GeV. The GBM Public GRB burst catalog\footnote{\url{https://heasarc.gsfc.nasa.gov/W3Browse/fermi/fermigbrst.html}} provides up-to-date information on all triggers classified as GRBs since the beginning of the mission. Considering the requirements of the signal-to-noise ratio (S/N) for time-resolved analysis, we just deal with the long GRBs that are bright enough to have an integral flux greater than $\mathrm{2 \, \times \, 10^{-5} \, erg \, cm^{-2}}$ over the burst duration. There are 386 GRBs satisfying such a brightness condition before January 2023. Furthermore, only those GRBs, whose lightcurve pulses are significantly identified and countable, are selected. Finally, 78 bright long GRBs are selected as our sample. \par

\subsection{Detector, Source and Background Selection}

There are 14 detectors on board Fermi/GBM, of which 12 Sodium Iodide (NaI, n0-nb) are used to detect photons in the energy range of 8\,keV-1\,MeV,  and 2 Bismuth Germanate (BGO, b0 and b1) are used to detect the photons ranged in 200\,keV-40\,MeV, respectively. The NaI detectors are distributed around the periphery of the spacecraft, with three detectors on each face to achieve full sky coverage. The BGO detectors are located on both sides of the spacecraft, which can similarly cover the entire sky.
In general, NaI detectors n0-n5 correspond to BGO detector b0, and NaI detectors n6-nb correspond to BGO detector b1. We selected data from three NaI detectors and one BGO detector at the optimal observation location.
\par

The data of GBM are stored in three types: the CTIME data, the CSPEC data, and the TTE (Time-Tagged Event) data. The CTIME data has a time resolution of 0.064\,s with 8 energy channels, and the CSPEC data has a time resolution of 1.024\,s with 128 energy channels. The TTE data has the same 128 energy channels as the CSPEC data and has a time resolution of 2\,$\upmu$s, containing both time and energy information for individual photons. Both CSPEC and TTE data have higher spectral resolution and can be used for spectral analysis. Because the TTE data has the highest temporal resolution, it is more suitable for time-resolved spectral analysis. The TTE and response files of the selected detector are used for spectral analysis in this work.
\par

The GRB source intervals we choose are generally slightly longer than the ${T_{90}}$ reported in the NASA/HEASARC database, as this would include all relevant features of the light curve. Meanwhile, we select the photons in the interval of tens of seconds before the trigger time and the interval of tens of seconds after the end of the emission to estimate the background. A polynomial of order 0 to 4 is applied to fit the background photon counts for all 128 energy channels and the optimal polynomial has been determined by a likelihood ratio test. The polynomial is then interpolated into the source interval to obtain the background photon count estimate. We have jointly fitted the spectral data of three NaI detectors and one BGO detector. The spectral energy ranges of the NaI detectors were set to 10-30\,keV and 40-900\,keV (excluding the k-edge\footnote{\url{https://fermi.gsfc.nasa.gov/ssc/data/analysis/caveats.html}} at 33.17\,keV), and the spectral energy ranges of the BGO detector was set to 250\,keV-40\,MeV.

\subsection{Lightcurve Time Binning}
The key premise for time-resolved spectrum analysis is to appropriately divide the lightcurve into time bins. An inappropriate division method may deviate from the real situation and thus lead to fuzzy results. There are three commonly used ways to divide time bins based on different considerations: uniform time bins, constant signal-to-noise ratio per bin, and Bayesian Blocks (BBlocks, see \cite{Scargle+etal+2013}). To minimize the variation in radiation over a time interval, which would mask the true spectral shape, BBlocks is thought to be the most appropriate method (\citealt{Burgess+etal+2014}). We first use the BBlocks method to divide the TTE lightcurve of the brightest NaI detector into time bins, with setting a false alarm probability of ${p_0 = 0.01}$ (\citealt{ Yu+etal+2019, Li+etal+2021}). And then, the time bins are transferred and used to other detectors. \par

The time bins divided by BBlocks have different S/N ratios. Sometimes the S/N in a certain time interval might be too low to satisfy the requirements of the statistical significance fitting. The statistical significance $S$ currently used is a test statistic that includes information on the signal-to-noise ratio, which is applicable to Poisson sources with a Gaussian background (\citealt{Vianello+etal+2018}). It is necessary to choose the time bins with ${S \geq 20}$ to ensure that there are enough photons to make the model parameters converge well (\citealt{Yu+etal+2019, Ryde+etal+2019, Li+etal+2021}). In addition, there are at least four-time bins  in each burst satisfying ${S \geq 20}$, so it is meaningful to study
\onecolumn
\begin{center}
\tablecaption{Basic Information of GRB Sample
\label{tab1: basic information}}
\small
\tablefirsthead{\hline\noalign{\smallskip} 
GRB   & Detectors & $T_{90}$   & $\Delta T_{\rm src}$ & $\Delta T_{\rm bkg,1}$ & $\Delta T_{\rm bkg,2}$ & $\rm Spectra(N_{S \geq 20})$ & Pulse \\ 
 & & ($\rm s$)   & ($\rm s$) & ($\rm s$)& ($\rm s$) & ($\rm N$) & ($\rm N$) \\ \hline\noalign{\smallskip}}
\tablehead{
\multicolumn{8}{l}{{Continued from previous page}} \\ 
\hline\noalign{\smallskip}
GRB   & Detectors & $T_{90}$   & $\Delta T_{\rm src}$ & $\Delta T_{\rm bkg,1}$ & $\Delta T_{\rm bkg,2}$ & $\rm Spectra(N_{S \geq 20})$ & Pulse \\ 
 & & ($\rm s$)   & ($\rm s$) & ($\rm s$)& ($\rm s$) & ($\rm N$) & ($\rm N$) \\
\hline\noalign{\smallskip}}
\tabletail{%
\multicolumn{8}{r}{{Continued on next page}} \\ \hline\noalign{\smallskip}}
\tablelasttail{\noalign{\smallskip}\hline}
\begin{supertabular}{cccccccc}
    081009140       & (n3),n4,n7,b1  & 41.345        & 0.0 to 55.0    & -25.0 to -10.0 & 60.0 to 80.0   & 28(20)         & 2 \\
    081125496       & n9,(na),nb,b1  & 9.280         & 0.0 to 12.0    & -25.0 to -10.0 & 20.0 to 40.0   & 9(6)           & 1 \\
    081215784       & n9,(na),nb,b1  & 5.568         & 0.0 to 10.0    & -25.0 to -10.0 & 15.0 to 30.0   & 26(22)         & 3 \\
    081221681       & n0,n1,(n2),b0  & 29.697        & 0.0 to 40.0    & -25.0 to -10.0 & 50.0 to 65.0   & 16(14)         & 2 \\
    081224887       & n6,n7,(n9),b1  & 16.448        & 0.0 to 25.0    & -25.0 to -10.0 & 30.0 to 60.0   & 10(7)          & 1 \\
    090719063       & n6,n7,(n8),b1  & 11.392        & -1.0 to 25.0   & -25.0 to -10.0 & 35.0 to 50.0   & 14(11)         & 1 \\
    090820027       & n1,(n2),n5,b0  & 12.416        & 29.0 to 50.0   & -25.0 to -10.0 & 65.0 to 80.0   & 21(18)         & 1 \\
    090902462       & n0,(n1),n2,b0  & 19.328        & 0.0 to 30.0    & -25.0 to -10.0 & 35.0 to 50.0   & 51(47)         & 3 \\
    090926181       & n3,n6,(n7),b1  & 13.760        & 0.0 to 20.0    & -25.0 to -10.0 & 30.0 to 45.0   & 26(24)         & 2 \\
    091127976       & (n6),n9,na,b1  & 8.701         & -1.0 to 10.0   & -25.0 to -10.0 & 20.0 to 40.0   & 22(17)         & 3 \\
    100324172      & n1,(n2),n5,b0  & 17.920        & 0.0 to 22.0    & -25.0 to -10.0 & 30.0 to 45.0   & 11(7)          & 2 \\
    100707032      & n4,n7,(n8),b1  & 81.793        & 0.0 to 83.0    & -25.0 to -10.0 & 100.0 to 115.0 & 16(12)         & 1 \\
    100719989      & n3,(n4),n5,b0  & 21.824        &  -1.0 to 25.0  & -25.0 to -10.0 & 40.0 to 60.0   & 19(12)         & 3 \\
    101123952      & n9,(na),nb,b1  & 103.938       & 35.0 to 160.0  & -25.0 to -10.0 & 175.0 to 200.0 & 48(33)         & 5 \\
    101126198      & (n7),n8,nb,b1  & 43.837        & 0.0 to 55.0    & -25.0 to -10.0 & 60.0 to 75.0   & 10(9)          & 1 \\
    110301214      & n7,(n8),nb,b1  & 5.693         & -1.0 to 10.0   & -25.0 to -10.0 & 20.0 to 40.0   & 18(14)         & 2 \\
    110625881      & n7,n8,(nb),b1  & 26.881        & 0.0 to 35.0    & -25.0 to -10.0 & 50.0 to 65.0   & 38(24)         & 3 \\
    110721200      & (n6),n7,n9,b1  & 21.822        & 0.0 to 25.0    & -25.0 to -10.0 & 35.0 to 50.0   & 9(8)           & 1 \\
    110920546      & (n0),n1,n3,b0  & 160.771       & -1.0 to 170.0  & -25.0 to -10.0 & 175.0 to 190.0 & 11(9)          & 1 \\
    111220486      & n0,(n1),n2,b0  & 39.041        & -8.0 to 40.0   & -25.0 to -10.0 & 65.0 to 80.0   & 35(20)         & 2 \\
    120119170      & n7,n9,(nb),b1  & 55.297        & 0.0 to 70.0    & -25.0 to -10.0 & 80.0 to 95.0   & 14(10)         & 1 \\
    120328268      & n7,n9,(nb),b1  & 29.697        & 0.0 to 50.0    & -25.0 to -10.0 & 75.0 to 90.0   & 23(20)         & 2 \\
    120711115      & (n2),n8,na,b1  & 44.033        & 0.0 to 120.0   & -25.0 to -10.0 & 145.0 to 160.0 & 33(25)         & 2 \\
    120728434      & n1,(n2),n5,b0  & 100.481       & 0.0 to 120.0   & -25.0 to -10.0 & 200.0 to 215.0 & 50(41)         & 5 \\
    120919309      & (n1),n2,n5,b0  & 21.248        & -2.0 to 35.0   & -25.0 to -10.0 & 45.0 to 60.0   & 9(5)           & 1 \\
    130518580      & (n3),n6,n7,b1  & 48.577        & 5.0 to 65.0    & -25.0 to -10.0 & 85.0 to 100.0  & 20(16)         & 1 \\
    130606497      & n7,(n8),nb,b1  & 52.225        & 0.0 to 63.0    & -25.0 to -10.0 & 75.0 to 95.0   & 43(36)         & 4 \\
    130704560      & n3,(n4),n5,b0  & 6.400         & -2.0 to 13.0   & -25.0 to -10.0 & 25.0 to 40.0   & 19(16)         & 3 \\
    131014215      & n9,na,(nb),b1  & 3.200         &  1.0 to 6.0    & -25.0 to -10.0 & 20.0 to 40.0   & 29(27)         & 2 \\
    140206275      & n0,(n1),n3,b0  & 146.690       & 0.0 to 50.0    & -25.0 to -10.0 & 70.0 to 90.0   & 29(17)         & 2 \\
    140213807      & n0,(n1),n2,b0  & 18.624        & 0.0 to 20.0    & -25.0 to -10.0 & 35.0 to 50.0   & 13(10)         & 2 \\
    140329295      & n7,(n8),nb,b1  & 21.248        & -1.0 to 30.0   & -25.0 to -10.0 & 40.0 to 65.0   & 26(18)         & 2 \\
    141028455      & (n6),n7,n9,b1  & 31.489        & 0.0 to 50.0    & -25.0 to -10.0 & 60.0 to 85.0   & 14(11)         & 1 \\
    150127589      & n7,n8,(nb),b1  & 60.929        & 0.0 to 75.0    & -25.0 to -10.0 & 100.0 to 125.0 & 23(17)         & 2 \\
    150201574      & (n3),n4,n7,b0  & 15.616        & 0.0 to 25.0    & -25.0 to -10.0 & 50.0 to 75.0   & 25(25)         & 2 \\
    150330828      & n1,(n2),n5,b0  & 153.859       & 0.0 to 170.0   & -25.0 to -10.0 & 225.0 to 240.0 & 41(34)         & 3 \\
    150403913      & n0,(n3),n4,b0  & 22.272        & 0.0 to 45.0    & -25.0 to -10.0 & 50.0 to 75.0   & 16(12)         & 2 \\
    150902733      & n0,n1,(n3),b0  & 13.568        & 0.0 to 25.0    & -25.0 to -10.0 & 50.0 to 75.0   & 16(12)         & 1 \\
    160113398      & n7,n8,(nb),b1  & 24.576        & 20.0 to 60.0   & -25.0 to -10.0 & 75.0 to 100.0  & 15(10)         & 1 \\
    160530667      & n1,(n2),n5,b0  & 9.024         & -2.0 to 20.0   & -25.0 to -10.0 & 40.0 to 60.0   & 22(19)         & 1 \\
    160905471      & (n6),n7,n9,b1  & 33.537        & 0.0 to 50.0    & -25.0 to -10.0 & 75.0 to 100.0  & 17(11)         & 2 \\
    160910722      & n1,n2,(n5),b0  & 24.320        & 0.0 to 40.0    & -25.0 to -10.0 & 60.0 to 85.0   & 20(18)         & 1 \\
    170405777      & n6,(n7),n9,b1  & 78.593        & 0.0 to 88.0    & -25.0 to -10.0 & 100.0 to 115.0 & 21(14)         & 3 \\
    170522657      & (n0),n1,n2,b0  & 7.424         & 0.0 to 10.0    & -25.0 to -10.0 & 15.0 to 30.0   & 10(6)          & 2 \\
    170808936      & n1,n3,(n5),b0  & 17.664        & 0.0 to 30.0    & -25.0 to -10.0 & 50.0 to 65.0   & 39(37)         & 3 \\
    170826819      & n9,na,(nb),b1  & 11.008        & 0.0 to 15.0    & -25.0 to -10.0 & 25.0 to 40.0   & 15(11)         & 3 \\
    171210493      & n0,(n1),n2,b0  & 143.107       & 0.0 to 175.0   & -25.0 to -10.0 & 185.0 to 200.0 & 14(9)          & 1 \\
    171227000      & n2,n4,(n5),b0  & 37.633        & 0.0 to 55.0    & -25.0 to -10.0 & 60.0 to 75.0   & 43(40)         & 3 \\
    180113418      & n1,(n2),n9,b0  & 24.576        & 2.0 to 32.0    & -25.0 to -10.0 & 40.0 to 55.0   & 22(21)         & 2 \\
    180305393      & n1,(n2),na,b0  & 13.056        & 0.0 to 20.0    & -25.0 to -10.0 & 35.0 to 50.0   & 12(10)         & 1 \\
    180720598      & n6,(n7),n8,b1  & 48.897        & 0.0 to 60.0    & -25.0 to -10.0 & 100.0 to 115.0 & 75(70)         & 5 \\
    180728728      & n3,n6,(n7),b1  & 6.400         & 9.0 to 21.0    & -25.0 to -10.0 & 35.0 to 50.0   & 22(18)         & 1 \\
    181227262      & n1,(n2),na,b1  & 13.184        & 0.0 to 20.0    & -25.0 to -10.0 & 35.0 to 50.0   & 25(22)         & 2 \\
    190114873      & n3,(n4),n8,b0  & 116.354       & 0.0 to 120.0   & -25.0 to -10.0 & 130.0 to 145.0 & 49(45)         & 3 \\
    190530430      & n0,(n1),n2,b0  & 18.432        & 0.0 to 25.0    & -25.0 to -10.0 & 35.0 to 50.0   & 54(54)         & 3 \\
    190720613      & n0,(n1),n2,b0  & 6.144         & -1.0 to 9.0    & -25.0 to -10.0 & 15.0 to 35.0   & 13(9)          & 3 \\
    190727846      & n0,n1,(n3),b0  & 35.073        & 0.0 to 41.0    & -25.0 to -10.0 & 50.0 to 75.0   & 30(19)         & 4 \\
    190731943      & n6,n7,(n9),b1  & 15.872        & 0.0 to 22.0    & -25.0 to -10.0 & 35.0 to 50.0   & 18(14)         & 1 \\
    200101861      & n1,n3,(n5),b0  & 9.984         & 0.0 to 20.0    & -25.0 to -10.0 & 30.0 to 45.0   & 25(19)         & 2 \\
    200125864      & n0,n1,(n2),b0  & 5.824         & -1.0 to 10.0   & -25.0 to -10.0 & 20.0 to 40.0   & 33(30)         & 5 \\
    200313071      & n4,n7,(n8),b1  & 13.568        & -1.0 to 15.0   & -25.0 to -10.0 & 25.0 to 40.0   & 16(12)         & 1 \\
    200412381      & n6,(n7),n8,b1  & 6.080         & 0.0 to 14.0    & -25.0 to -10.0 & 25.0 to 40.0   & 30(25)         & 2 \\
    200826923      & n1,n2,(n5),b0  & 7.424         & 0.0 to 11.0    & -25.0 to -10.0 & 20.0 to 35.0   & 18(15)         & 1 \\
    200829582      & n4,n6,(n8),b1  & 6.912         & 15.0 to 30.0   & -25.0 to -10.0 & 40.0 to 55.0   & 18(15)         & 1 \\
    201016019      & n3,n4,(n5),b0  & 2.944         & 0.0 to 11.0    & -25.0 to -10.0 & 20.0 to 35.0   & 27(18)         & 1 \\
    201216963      & n9,(na),nb,b1  & 29.953        & 0.0 to 40.0    & -25.0 to -10.0 & 55.0 to 70.0   & 33(31)         & 2 \\
    210406949      & n1,n3,(n5),b0  & 19.712        & 0.0 to 30.0    & -25.0 to -10.0 & 45.0 to 60.0   & 17(16)         & 3 \\
    210610827      & (n9),na,nb,b1  & 55.041        & 10.0 to 90.0   & -25.0 to -10.0 & 120.0 to 135.0 & 17(14)         & 3 \\
    210714331      & n9,na,(nb),b1  & 42.369        & 0.0 to 45.0    & -25.0 to -10.0 & 55.0 to 70.0   & 15(11)         & 1 \\
    210801581      & n9,(na),nb,b1  & 13.824        & -1.0 to 17.0   & -25.0 to -10.0 & 25.0 to 40.0   & 14(10)         & 2 \\
    211019250      & n0,(n1),n2,b0  & 47.361        & 0.0 to 52.0    & -25.0 to -10.0 & 60.0 to 75.0   & 15(10)         & 1 \\
    220304228      & n8,(na),nb,b1  & 31.489        & 0.0 to 40.0    & -25.0 to -10.0 & 55.0 to 70.0   & 11(8)          & 1 \\
    220426285      & n0,n1,(n2),b0  & 5.632         & 0.0 to 10.0    & -25.0 to -10.0 & 20.0 to 35.0   & 24(23)         & 2 \\
    220527387      & n6,(n7),n8,b1  & 10.496        & 0.0 to 15.0    & -25.0 to -10.0 & 25.0 to 40.0   & 23(20)         & 4 \\
    220910242      & n7,n8,(nb),b1  & 4.224         & 0.0 to 11.0    & -25.0 to -10.0 & 20.0 to 35.0   & 23(18)         & 3 \\
    221022955      & n3,(n4),n5,b0  & 31.744        & 0.0 to 50.0    & -25.0 to -10.0 & 60.0 to 75.0   & 20(17)         & 2 \\
    221023862      & n0,(n1),n2,b0  & 39.169        & 0.0 to 50.0    & -25.0 to -10.0 & 60.0 to 75.0   & 32(26)         & 1 \\
    221209243      & n1,(n2),n5,b0  & 4.160         & 0.0 to 8.0     & -25.0 to -10.0 & 15.0 to 35.0   & 15(11)         & 1 \\
\end{supertabular}
\tablecomments{0.86\textwidth}{Fermi/GBM burst ID (column 1), utilized detector with the brightest response indicated in parentheses (column 2), duration represented by $T_{90}$ (column 3), source intervals $\Delta T_{\rm src}$ (columns 4), and background intervals $\Delta T_{\rm bkg,1}$ before the trigger time (columns 5), and background intervals $\Delta T_{\rm bkg,2}$ after the end of emission (columns 6), number of BBlock time bins (column 7), number of time bins with statistical significance greater than or equal to 20 enclosed in parentheses within column 7, and count of pulses observed in each burst (column 8).}
\end{center}
\twocolumn
\noindent the spectral evolution with sufficient time-resolved spectra. \par

The sample yielded a total of 1814 time-resolved spectra, of which 1490 spectra met the statistical significance of ${S \geq 20}$. The basic information of our sample is listed in Table~\ref{tab1: basic information}, including the Fermi/GBM burst ID (column 1), the detector used (column 2), the duration $T_{90}$ (column 3), the selected source and background intervals (columns 4 to 6), the number of BBlocks time bins (column 7), and the number of pulses per burst (column 8).

\subsection{Spectral Fitting}
The typical model  used in this work to fit the GRB spectra is the Band function \citep{Band+etal+1993} which is described as follows:
\begin{equation}\label{eq1}
\begin{array}{l}
N_{\mathrm{Band}}(E) =A\times\\
\left\{\begin{array}{ll}
\left(\frac{E}{E_{\mathrm{piv}}}\right)^{\alpha} e^{-E / E_{0}}, & (\alpha-\beta) E_{0} \geqslant E \\
\left(\frac{(\alpha-\beta) E_{0}}{E_{\mathrm{piv}}}\right)^{(\alpha-\beta)} e^{(\beta-\alpha)}\left(\frac{E}{E_{\mathrm{piv}}}\right)^{\beta}, & (\alpha-\beta) E_{0} \leqslant E
\end{array},\right.
\end{array}
\end{equation}
where $A$ is the normalization factor in unit of $\rm photon\,cm^{-2}\,keV^{-1}s^{-1}$; $E_{\mathrm{piv}}$ is the pivot energy ﬁxed at 100\,keV; $\alpha$ and $\beta$ are the low-energy and high-energy photon spectral indices, respectively; $E_0$ is break energy of the spectrum. The peak energy $E_{\rm p}$ of the $\nu F_{\nu}$ spectrum is related to $E_0$ through $E_{\rm p}\,=\,(2+\alpha)E_0$. \par

The Multi-Mission Maximum Likelihood Framework package (3ML\footnote{\url{https://threeml.readthedocs.io/en/stable/index.html}}, \citealt{Vianello+etal+2015}) is used for spectral fitting and parameter estimation in this work. There are two approaches available in 3ML for fitting data and models: the maximum likelihood estimation (MLE) method and the Bayesian method. In this study, we employ Bayesian parameter estimation, which involves accessing the posterior distribution of model parameters using a specific sampling algorithm. The optimal model parameters are estimated from the posterior probability distribution, which is evaluated by combining the prior distribution with the likelihood function that quantifies how well the model matches the observed data. Typical spectral parameters from the previous Fermi/GBM catalog are selected as priors, where the normalization factor ($A$) is a logarithmic uniform distribution (logU), the low and high energy indices $\alpha$ and $\beta$ are Gaussian (G), and the peak energy $E_{\rm p}$ of the $\nu F_{\nu}$ spectrum is a logarithmic normal distribution (logN). We apply $emcee$ (\citealt{Foreman+etal+2013}) in this work for posterior sampling, which is a Markov chain Monte Carlo (MCMC) sampling algorithm. We specify the number of walkers to 20, and the number of global samples to 10,000. During MCMC sampling, it requires a certain number of samples for the Markov chain to reach convergence and achieve a steady-state of the parameter distribution. Therefore, the samples that did not reach the steady-state distribution in the previous period should be discarded. ln this study, we excluded the initial 25\% of 10,000 MCMC samples for each parameter sampling and only considered the last 75\% as representative samples from the posterior distribution that reached convergence.\par

\textit{}There are two main aspects of spectral fitting. The first one is parameter estimation, which aims to determine the best-fitting parameters and their uncertainties for a given model. The second aspect involves evaluating the concordance between the model and the data, commonly referred to as the goodness-of-fit (GOF). Traditionally, the GOF is evaluated using a reduced chi-square ($\chi^{2}$), denoted as $\chi^{2}/\mathrm{dof}$, where $\mathrm{dof}$ represents the degree of freedom. A good fit is usually indicated by a value close to 1. However, this approach may lack reliability and accuracy when the data deviates from a normal distribution. In the case of MLE or Bayesian methods, alternative statistical measures can be used to assess the goodness-of-fit of the data. For example, Cstat (\citealt{Cash+etal+1979}) and Pgstat\footnote{\url{https://heasarc.gsfc.nasa.gov/xanadu/xspec/manual/XSappendixStatistics.html}}, where $\mathrm{Pgstat\,= \, -2ln}L(\theta)$, with $L(\theta)$ representing the likelihood value as a function of the free parameter $\theta$. The minimum of the $\mathrm{-ln}L(\theta)$ function corresponds to the maximum likelihood. Different statistics are utilized in Fermi data analysis due to the small counts in each time bin, which renders it inappropriate to treat each bin as a single observation from a normal distribution. Instead, a Poisson distribution is more suitable. It is widely acknowledged that as the bin counts increase sufficiently, the Poisson distribution approximates a normal distribution. At this point, Pgstat is expected to closely approximate $\chi^2$, and it can be anticipated that $\mathrm{-2ln}L(\theta)/\mathrm{dof}$ will approach 1. However, if the bin counts are insufficiently large, there is no basis for expecting $\mathrm{-2ln}L(\theta)/\mathrm{dof}$ equal to 1. \par

\section{Spectral Analysis Results}

We employed the Bayesian approach (emcee) to conduct spectral fitting on 1490 time-resolved spectra with a significance of ${S \geq 20}$ obtained from 78 bright long GRBs. As an illustrative example, Table \ref{tab2:fit results} presents the spectral fitting results for GRB 171227000. Similar to the method employed by \cite{Li+etal+2019+apjs.b}, we utilized Pgstat as the statistical measure instead of $\chi^2$. Consequently, $\mathrm{-2ln}L(\theta)/\mathrm{dof}$ does not typically converge to unity. As previously elucidated, this occurrence is not uncommon due to insufficient photon counts in each bin.

\begin{table*}
\bc
\caption{Time-resolved Spectral Fit Results of GRB 171227000}
\label{tab2:fit results}
\small
 \begin{tabular}{cccccccc}
  \hline\noalign{\smallskip}
    t$_{1}$$\sim$t$_{2}$& $\rm S$ &  $\alpha$ & $\beta$ & $\rm E_{p}$ & $\rm F$ & BIC/-$\ln$(posterior)       & dof\\
    $\rm (s)\sim(s)$& & & &$\rm (keV)$&$\rm (erg\,cm^{-2}\,s^{-1})$& \\
  \hline\noalign{\smallskip}
0.0$\sim$6.348&22.23&-0.16$^{+0.11}_{-0.11}$&-2.36$^{+0.13}_{-0.33}$&467.32$^{+55.76}_{-36.85}$&3.07$^{+0.56}_{-0.43}$$\times$10$^{-6}$&5197.8 / 2586.63&463\\
6.348$\sim$14.39&46.21&-0.43$^{+0.06}_{-0.05}$&-2.81$^{+0.22}_{-0.32}$&411.03$^{+21.77}_{-23.85}$&3.3$^{+0.33}_{-0.26}$$\times$10$^{-6}$&5590.1 / 2782.75&463\\
14.39$\sim$15.099&23.39&-0.46$^{+0.09}_{-0.09}$&-2.42$^{+0.15}_{-0.33}$&562.44$^{+76.35}_{-56.48}$&8.86$^{+1.46}_{-1.21}$$\times$10$^{-6}$&2232.5 / 1103.95&438\\
15.099$\sim$15.548&28.2&-0.49$^{+0.08}_{-0.08}$&-2.57$^{+0.16}_{-0.34}$&627.34$^{+77.33}_{-59.11}$&13.65$^{+1.81}_{-1.47}$$\times$10$^{-6}$&1783.2 / 879.3&398\\
15.548$\sim$16.821&68.64&-0.59$^{+0.03}_{-0.03}$&-2.68$^{+0.08}_{-0.21}$&1097.97$^{+89.72}_{-51.14}$&32.89$^{+1.46}_{-1.51}$$\times$10$^{-6}$&3400.1 / 1687.77&463\\
16.821$\sim$17.341&59.64&-0.62$^{+0.06}_{-0.03}$&-2.7$^{+0.12}_{-0.39}$&1284.91$^{+94.7}_{-155.36}$&51.51$^{+3.05}_{-4.02}$$\times$10$^{-6}$&2197.9 / 1086.64&413\\
17.341$\sim$17.648&36.2&-0.66$^{+0.05}_{-0.06}$&-2.48$^{+0.11}_{-0.3}$&891.03$^{+132.61}_{-82.9}$&28.37$^{+3.07}_{-2.57}$$\times$10$^{-6}$&1461.1 / 718.25&375\\
17.648$\sim$17.82&44.69&-0.46$^{+0.04}_{-0.06}$&-2.34$^{+0.08}_{-0.15}$&995.8$^{+110.72}_{-67.25}$&77.33$^{+5.72}_{-5.59}$$\times$10$^{-6}$&1020.7 / 498.05&313\\
17.82$\sim$18.185&47.03&-0.47$^{+0.05}_{-0.06}$&-2.37$^{+0.09}_{-0.17}$&660.95$^{+65.57}_{-40.52}$&34.54$^{+3.05}_{-2.84}$$\times$10$^{-6}$&1736.0 / 855.72&383\\
18.185$\sim$18.647&83.44&-0.6$^{+0.03}_{-0.03}$&-2.39$^{+0.07}_{-0.1}$&1446.55$^{+107.74}_{-89.14}$&114.67$^{+4.16}_{-4.26}$$\times$10$^{-6}$&2298.4 / 1136.91&400\\
18.647$\sim$20.433&170.86&-0.61$^{+0.01}_{-0.02}$&-2.55$^{+0.04}_{-0.06}$&1249.66$^{+44.03}_{-30.5}$&97.96$^{+1.8}_{-1.87}$$\times$10$^{-6}$&4366.1 / 2170.77&463\\
20.433$\sim$20.836&74.95&-0.64$^{+0.03}_{-0.03}$&-2.52$^{+0.09}_{-0.15}$&868.01$^{+66.73}_{-52.57}$&58.35$^{+3.16}_{-2.82}$$\times$10$^{-6}$&2006.2 / 990.81&387\\
20.836$\sim$21.554&73.28&-0.66$^{+0.02}_{-0.03}$&-2.68$^{+0.11}_{-0.19}$&799.44$^{+50.31}_{-35.94}$&33.31$^{+1.94}_{-1.88}$$\times$10$^{-6}$&2643.8 / 1309.6&438\\
21.554$\sim$22.212&81.8&-0.65$^{+0.03}_{-0.03}$&-2.73$^{+0.09}_{-0.23}$&854.0$^{+56.3}_{-41.49}$&41.64$^{+2.11}_{-2.09}$$\times$10$^{-6}$&2589.3 / 1282.35&436\\
22.212$\sim$22.728&65.1&-0.56$^{+0.05}_{-0.05}$&-2.26$^{+0.08}_{-0.11}$&575.6$^{+48.11}_{-40.39}$&36.1$^{+2.85}_{-2.62}$$\times$10$^{-6}$&2157.6 / 1066.51&413\\
22.728$\sim$22.985&35.72&-0.65$^{+0.07}_{-0.06}$&-2.7$^{+0.17}_{-0.36}$&502.48$^{+54.01}_{-43.21}$&15.35$^{+1.85}_{-1.46}$$\times$10$^{-6}$&1152.8 / 564.13&354\\
22.985$\sim$24.215&55.76&-0.65$^{+0.04}_{-0.05}$&-2.35$^{+0.1}_{-0.2}$&433.86$^{+43.5}_{-27.82}$&12.53$^{+1.37}_{-1.2}$$\times$10$^{-6}$&3101.5 / 1538.44&463\\
24.215$\sim$25.597&42.21&-0.62$^{+0.06}_{-0.05}$&-2.67$^{+0.17}_{-0.35}$&372.43$^{+27.83}_{-25.68}$&6.22$^{+0.7}_{-0.57}$$\times$10$^{-6}$&3217.4 / 1596.39&463\\
25.597$\sim$25.942&31.04&-0.69$^{+0.06}_{-0.08}$&-2.52$^{+0.15}_{-0.43}$&562.03$^{+96.66}_{-53.87}$&13.29$^{+2.0}_{-1.52}$$\times$10$^{-6}$&1525.4 / 750.43&380\\
25.942$\sim$26.489&27.53&-0.75$^{+0.11}_{-0.11}$&-2.02$^{+0.1}_{-0.27}$&376.37$^{+109.98}_{-61.92}$&9.99$^{+2.42}_{-1.87}$$\times$10$^{-6}$&1881.6 / 928.5&419\\
26.489$\sim$26.738&40.58&-0.71$^{+0.07}_{-0.05}$&-2.6$^{+0.23}_{-0.28}$&827.99$^{+162.28}_{-57.12}$&29.84$^{+4.33}_{-2.83}$$\times$10$^{-6}$&1270.9 / 623.15&351\\
26.738$\sim$27.252&40.04&-0.72$^{+0.06}_{-0.05}$&-2.69$^{+0.2}_{-0.33}$&521.31$^{+52.2}_{-44.9}$&12.71$^{+1.49}_{-1.1}$$\times$10$^{-6}$&1957.4 / 966.39&413\\
27.252$\sim$27.403&31.79&-0.54$^{+0.07}_{-0.15}$&-2.2$^{+0.07}_{-0.47}$&448.12$^{+140.82}_{-37.25}$&24.59$^{+4.8}_{-3.62}$$\times$10$^{-6}$&743.7 / 359.56&292\\
27.403$\sim$27.557&52.74&-0.63$^{+0.04}_{-0.06}$&-2.28$^{+0.08}_{-0.17}$&901.06$^{+144.15}_{-68.54}$&81.07$^{+6.48}_{-5.84}$$\times$10$^{-6}$&998.2 / 486.78&294\\
27.557$\sim$28.019&59.26&-0.71$^{+0.05}_{-0.04}$&-2.68$^{+0.13}_{-0.39}$&554.21$^{+69.06}_{-28.54}$&23.38$^{+2.24}_{-1.68}$$\times$10$^{-6}$&1967.8 / 971.6&402\\
28.019$\sim$28.712&63.39&-0.77$^{+0.04}_{-0.05}$&-2.43$^{+0.12}_{-0.26}$&506.63$^{+51.33}_{-40.18}$&19.37$^{+2.01}_{-1.74}$$\times$10$^{-6}$&2469.4 / 1222.41&437\\
28.712$\sim$29.102&37.0&-0.77$^{+0.08}_{-0.07}$&-2.56$^{+0.2}_{-0.35}$&452.41$^{+59.82}_{-58.44}$&12.27$^{+1.7}_{-1.32}$$\times$10$^{-6}$&1621.9 / 798.65&385\\
29.102$\sim$29.373&45.47&-0.69$^{+0.05}_{-0.06}$&-2.62$^{+0.18}_{-0.28}$&516.55$^{+55.31}_{-42.72}$&22.5$^{+2.41}_{-1.98}$$\times$10$^{-6}$&1380.0 / 677.73&360\\
29.373$\sim$29.91&43.72&-0.72$^{+0.05}_{-0.12}$&-2.36$^{+0.1}_{-0.51}$&340.43$^{+82.06}_{-25.01}$&11.37$^{+2.3}_{-1.45}$$\times$10$^{-6}$&1981.9 / 978.66&418\\
29.91$\sim$31.369&40.43&-0.59$^{+0.09}_{-0.11}$&-2.14$^{+0.08}_{-0.16}$&201.05$^{+30.1}_{-17.61}$&5.83$^{+0.92}_{-0.83}$$\times$10$^{-6}$&3138.9 / 1557.15&463\\
31.369$\sim$31.757&28.09&-0.79$^{+0.11}_{-0.09}$&-2.46$^{+0.2}_{-0.42}$&307.15$^{+50.88}_{-40.8}$&6.19$^{+1.38}_{-0.84}$$\times$10$^{-6}$&1501.2 / 738.28&384\\
31.757$\sim$32.395&22.1&-0.67$^{+0.15}_{-0.12}$&-2.26$^{+0.15}_{-0.41}$&196.11$^{+31.61}_{-22.8}$&3.38$^{+1.06}_{-0.68}$$\times$10$^{-6}$&1976.9 / 976.17&436\\
32.395$\sim$38.267&42.74&-0.86$^{+0.09}_{-0.09}$&-2.43$^{+0.13}_{-0.3}$&134.6$^{+15.93}_{-10.0}$&1.5$^{+0.23}_{-0.17}$$\times$10$^{-6}$&5008.1 / 2491.77&463\\
38.511$\sim$39.062&36.6&-0.83$^{+0.09}_{-0.09}$&-2.43$^{+0.18}_{-0.42}$&201.87$^{+30.08}_{-22.36}$&5.37$^{+1.12}_{-0.7}$$\times$10$^{-6}$&1887.5 / 931.47&421\\
39.062$\sim$40.274&44.06&-0.81$^{+0.14}_{-0.1}$&-2.07$^{+0.09}_{-0.12}$&149.42$^{+20.79}_{-22.83}$&5.86$^{+1.03}_{-0.9}$$\times$10$^{-6}$&2890.1 / 1432.77&463\\
40.274$\sim$42.928&39.29&-0.93$^{+0.14}_{-0.1}$&-2.17$^{+0.09}_{-0.19}$&121.26$^{+18.31}_{-15.46}$&2.44$^{+0.44}_{-0.38}$$\times$10$^{-6}$&3886.1 / 1930.74&463\\
42.928$\sim$45.357&26.46&-0.86$^{+0.24}_{-0.15}$&-2.25$^{+0.09}_{-0.24}$&91.84$^{+14.61}_{-13.73}$&1.4$^{+0.26}_{-0.22}$$\times$10$^{-6}$&3710.2 / 1842.81&463\\
45.357$\sim$46.238&27.02&-0.94$^{+0.14}_{-0.11}$&-2.75$^{+0.21}_{-0.38}$&117.4$^{+12.43}_{-11.17}$&1.72$^{+0.22}_{-0.15}$$\times$10$^{-6}$&2390.7 / 1183.07&449\\
46.238$\sim$49.915&31.2&-1.0$^{+0.23}_{-0.1}$&-2.63$^{+0.21}_{-0.31}$&83.84$^{+8.34}_{-11.93}$&0.94$^{+0.13}_{-0.08}$$\times$10$^{-6}$&4302.0 / 2138.72&463\\
49.915$\sim$51.705&29.18&-1.0$^{+0.19}_{-0.11}$&-2.71$^{+0.19}_{-0.35}$&78.4$^{+7.26}_{-7.79}$&1.08$^{+0.13}_{-0.09}$$\times$10$^{-6}$&3268.7 / 1622.03&463\\
  \noalign{\smallskip}\hline
\end{tabular}
\ec
\tablecomments{0.86\textwidth}{The start and stop times of the BBlock time bins (column 1), the significance $\rm S$ of time bins (column 2), the optimal parameters for the Band model (column 3-5), the derived energy flux (column 6), the Bayesian information criterion BIC and the $\rm -\ln(posterior)$ values (consistent with the $\mathrm{-ln}L(\theta)$ values) (column 7), as well as degrees of freedom (column 8).}
\end{table*}

\subsection{Parameter Distribution}
\begin{figure*}[!htbp]
	\centering
	\includegraphics[width=0.48\textwidth]{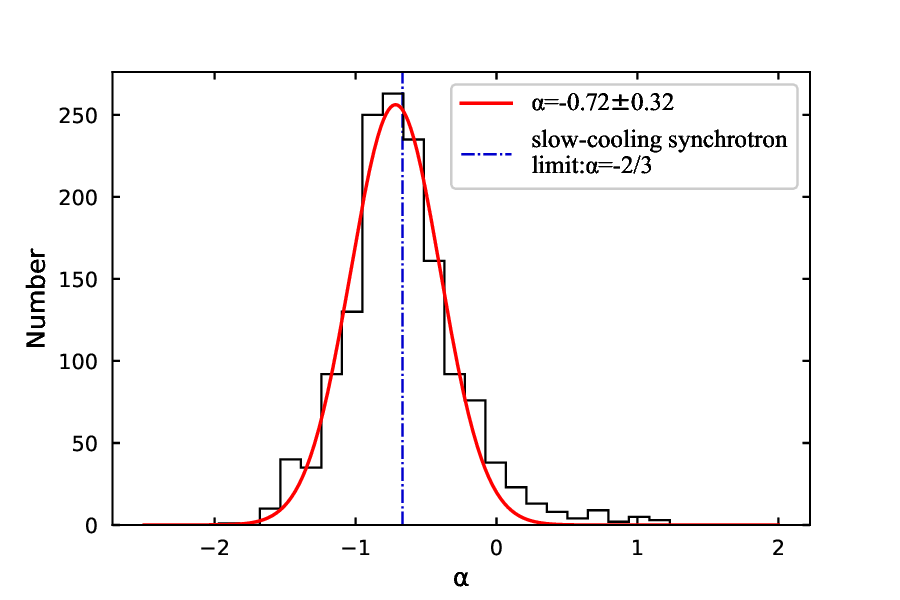}
	\includegraphics[width=0.48\textwidth]{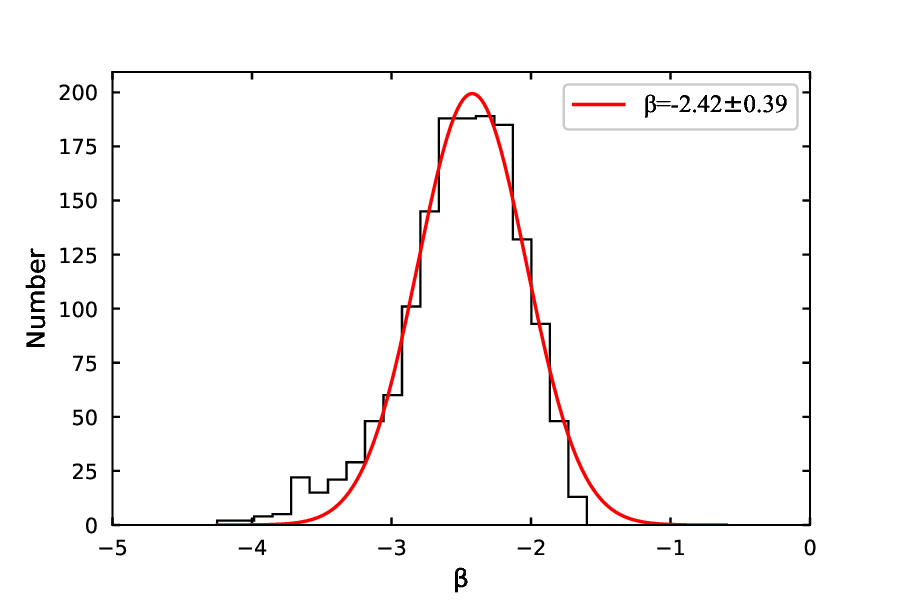}
    \includegraphics[width=0.48\textwidth]{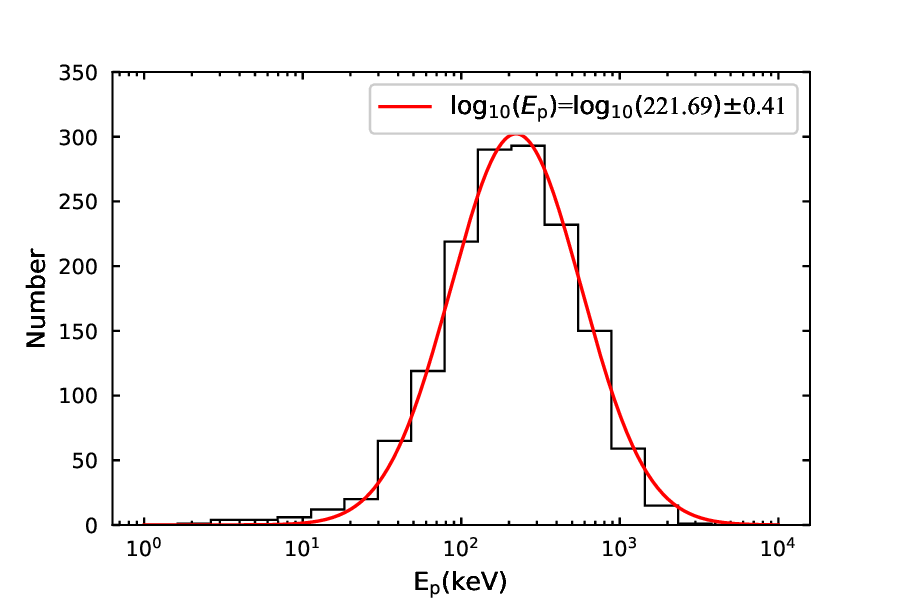}
	\includegraphics[width=0.48\textwidth]{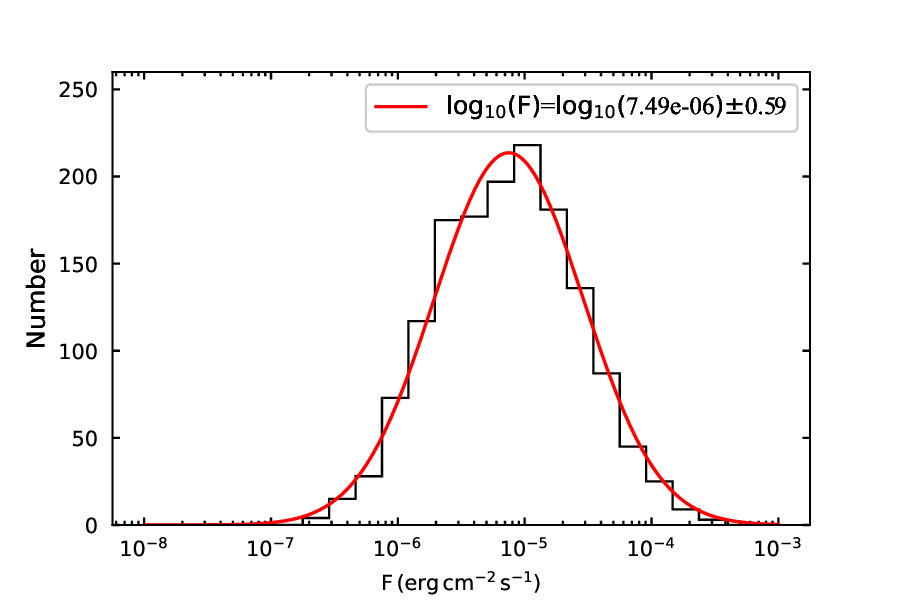}
	\caption{The distributions of the time-resolved spectral parameters $\alpha$ (upper left panel), $\beta$ (upper right panel), $E_{\rm p}$ (lower left panel), and $F$ (lower right panel). The low-energy index ($\alpha$) and high-energy index ($\beta$) are fitted with the Gaussian profile, and the peak energy ($E_{\rm p}$) and flux ($F$) are fitted using a log-normal distribution. The red line represents the fitted line of the parameter distribution histogram.}
	\label{fig1:global distribution}
\end{figure*}

The global parameter distribution, including $\alpha$, $\beta$, $E_{\rm p}$ and the derived parameter flux ($F$), $F$ is the best-fit model integrand in the energy range of 1 keV to 40 MeV, is depicted in Figure \ref{fig1:global distribution}. The Gaussian profiles were employed to fit the $\alpha$ and $\beta$ distributions (${\mathcal{N}}=\mu\pm\sigma$, where $\mu$ represents the average value and $\sigma$ denotes the corresponding standard deviation). On the other hand, due to their approximate lognormal, a lognormal profile was used to fit the distributions of $E_{\rm p}$ and $F$. The overall sample exhibited mean values and standard deviations as follows: $\alpha=-0.72 \pm 0.32$, $\beta=-2.42 \pm 0.39$, $\mathrm{log_{10}}{(E_{\rm p}\rm{/keV})}=\mathrm{log_{10}{(221.69)} \pm 0.41}$, $\mathrm{log_{10}}{(F\rm{/(erg\,cm^{-2}\,s^{-1}})} = \mathrm{log_{10}{(7.49e-6)} \pm 0.59}$. These findings are consistent with the results of previous studies(\citealt{Poolakkil+2021+ApJ}).\par

We conducted an analysis of 1490 spectra, where 669 spectra (44.9\%) of them displayed $\alpha$ values surpassing the synchrotron radiation limit ($\alpha > -2/3$). According to \cite{Acuner+etal+2020}, spectra with $\alpha > -0.5$ indicate a preference for the photosphere model. Therefore, based on their criteria, we identified 410 spectra (27.5\%) that exhibit a stronger inclination towards photosphere emission.\par
\begin{figure}
   \centering
  \includegraphics[width=0.48\textwidth, angle=0]{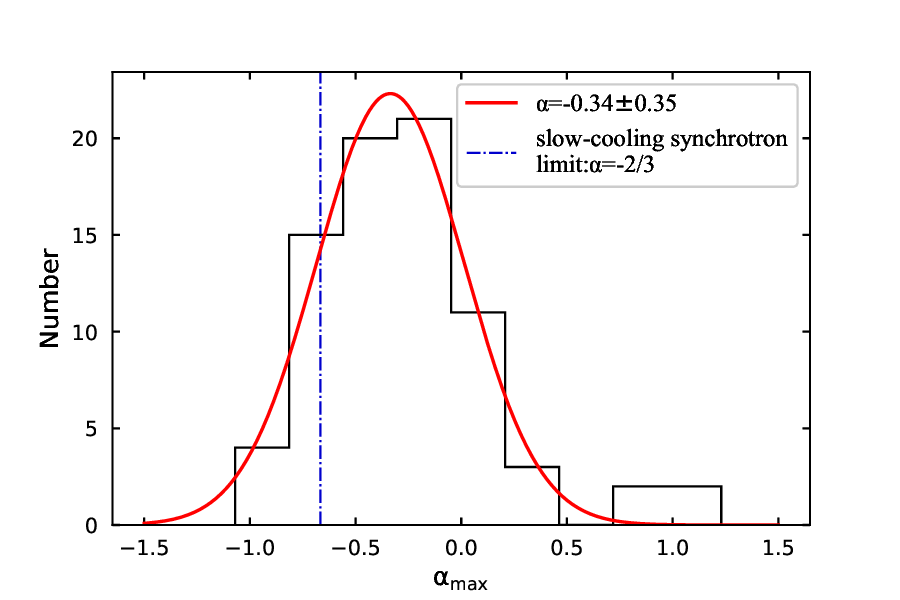}
   \caption{The distribution of the hardest low-energy index ($\alpha_{\rm max}$) in each GRB, and the red line represents a Gaussian fit to the $\alpha_{\rm max}$  distribution.}
     \label{fig2:alpha_max}
\end{figure}

Subsequently, the hardest low-energy index $\alpha_{\rm max}$ was identified for each burst, following a distribution as depicted in Figure \ref{fig2:alpha_max}, $\alpha_{\rm max}\, = \,-0.34 \pm 0.35$. Notably, 67 bursts (86\%) exhibited an $\alpha_{\rm max}$ value surpassing the synchrotron radiation limit, which significantly exceeded the proportion observed in the overall time-resolved spectrum. According to the criteria of \cite{Acuner+etal+2020}, 56 bursts (71.7\%) have $\alpha_{\rm max} > -0.5$, a pronounced inclination towards photosphere emission within this spectral range. \par

The presence of a harder low-energy index suggests that the simple synchrotron model is inadequate to fit the spectrum, and it might be usually considered as an indicator of the photosphere model. If a harder $\alpha$ is owed to the photosphere emission, a thermal component can be expected in the spectrum. However, as referred in Section \ref{sect:intro}, an inference of this kind could not so indubitably be drawn, considering the different details of the radiation process (e.g., \citealt{Lundman+etal+2013, Burgess+etal+2020}). Indeed, it might be misleading to judge the radiation mechanism with relying solely on whether the low-energy index exceeds the synchrotron radiation limit. A more reasonable method is to fit the GRB data directly using the physical models rather than empirical functions and compare the goodness of fit (e.g., \citealt{meng+etal+2018}). For example, by directly fitting the radiation model involving time-dependent cooling of synchrotron electrons to the observed data of GRBs, \cite{ Burgess+etal+2020} pointed out that the synchrotron emission call still be suitable for many GRBs, despite their $\alpha$ larger than $-2/3$. Some other works also prove the necessity and feasibility of implementing the directly fitting to data with physical model (\citealt{Zhang+etal+2016, Yang+etal+2023}), but it is beyond the scope of this work. \par

\subsection{Spectral Evolution}
 Previous studies have primarily focused on the evolution of parameters, particularly the peak energy $E_{\rm p}$, which has been extensively investigated in early research. The evolution of $E_{\rm p}$ exhibits various patterns: (i) the hard-to-soft pattern, (ii) the intensity-tracking pattern, and (iii) the soft-to-hard pattern or chaotic evolution. Subsequent studies have highlighted the prevalence of the first two patterns. In the case of multi-pulse GRBs, however, spectral evolution patterns become more intricate, and recent research has revealed a higher occurrence of rough tracking rather than smooth tracking (\citealt{Duan+etal+2020, Li+etal+2021}). Moreover, there is also noticeable evolution behavior in the low-energy index $\alpha$, which is more complex compared to the evolution of $E_{\rm p}$. \cite{Li+etal+2021} found that half of their samples exhibited tracking-patterns of $\alpha$ evolution. On the other hand, the high-energy index $\beta$ does not display a clear population trend, and its greater chaos-like behavior has not been statistically analyzed before.\par

 \begin{figure*}
	\centering
 \begin{minipage}{0.49\textwidth}
          \begin{minipage}{0.6\textwidth}
	\includegraphics[width=\textwidth]{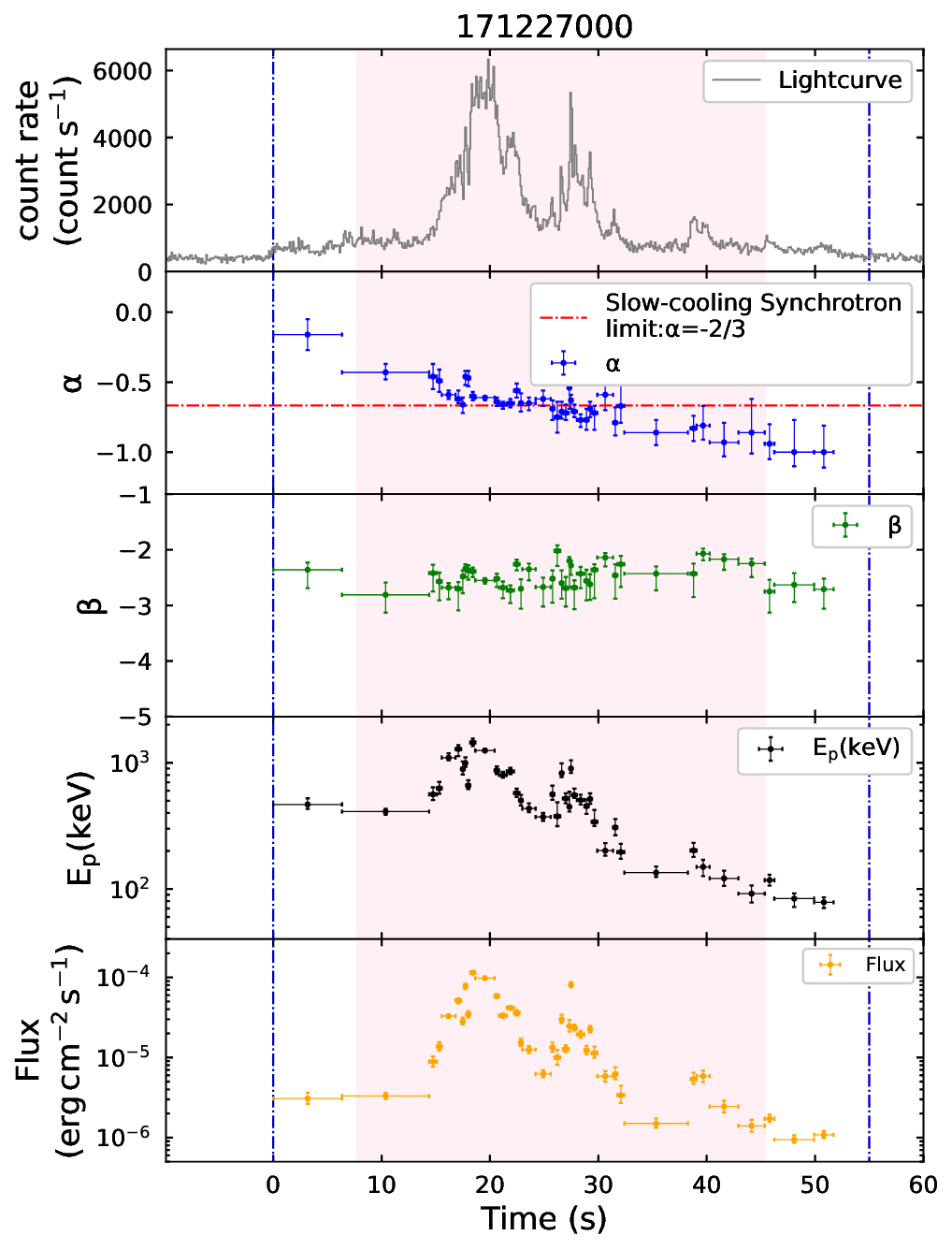}
         \end{minipage}
         \begin{minipage}{0.38\textwidth}
     \includegraphics[width=\textwidth]{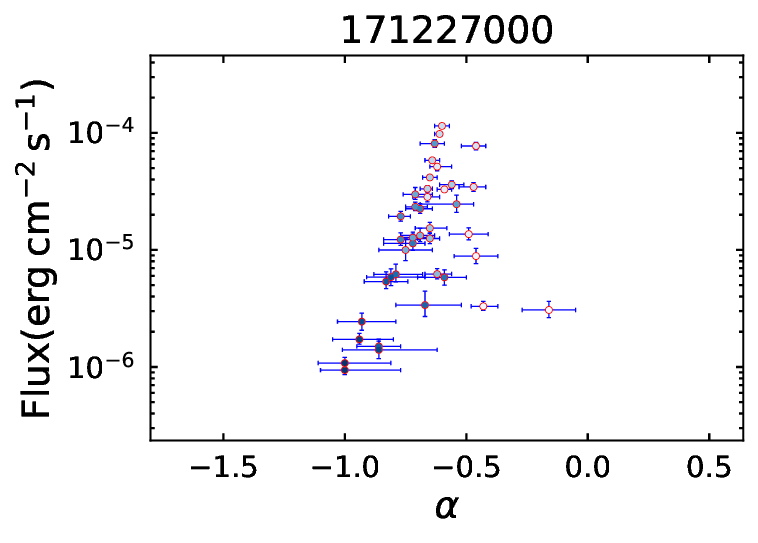}
     \includegraphics[width=\textwidth]{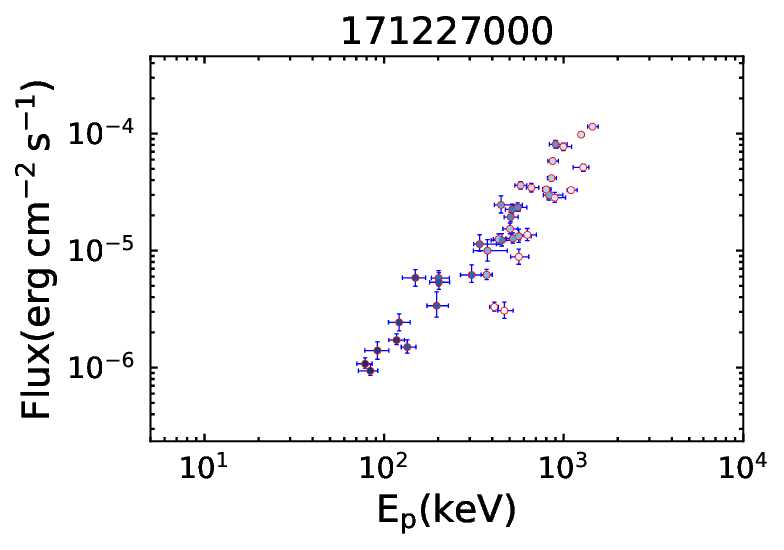}
     \includegraphics[width=\textwidth]{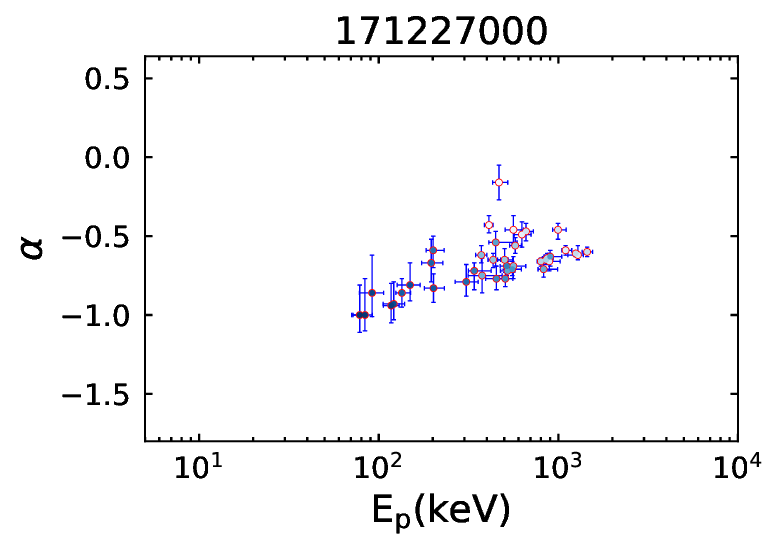}
     \end{minipage}
     \end{minipage} 
      \begin{minipage}{0.49\textwidth}
          \begin{minipage}{0.6\textwidth}
	\includegraphics[width=\textwidth]{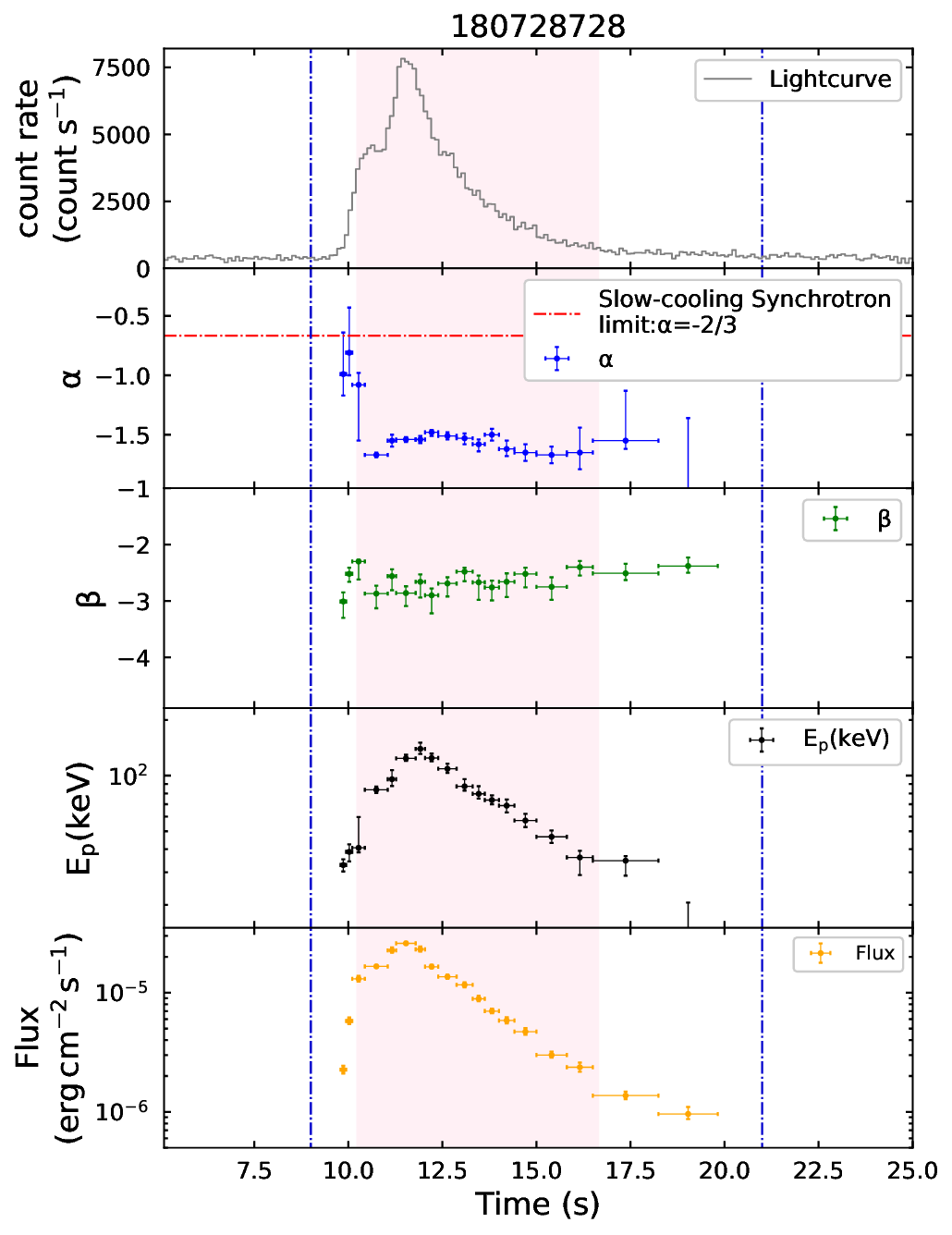}
         \end{minipage}
         \begin{minipage}{0.38\textwidth}
     \includegraphics[width=\textwidth]{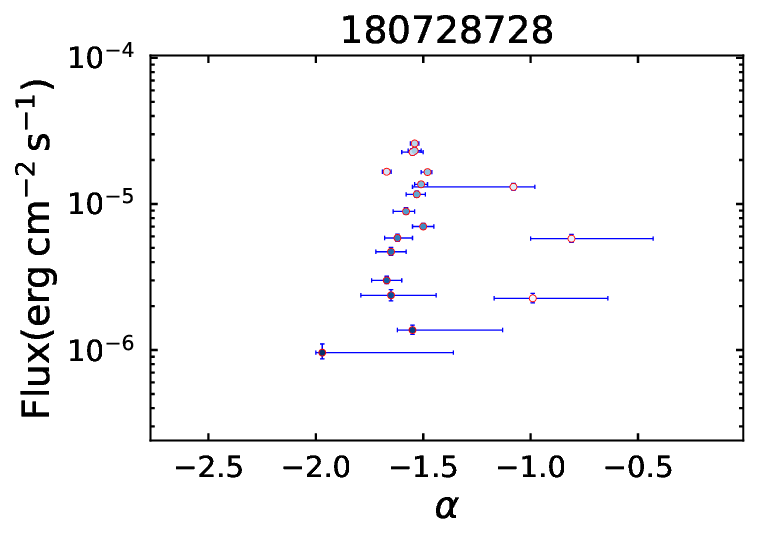}
     \includegraphics[width=\textwidth]{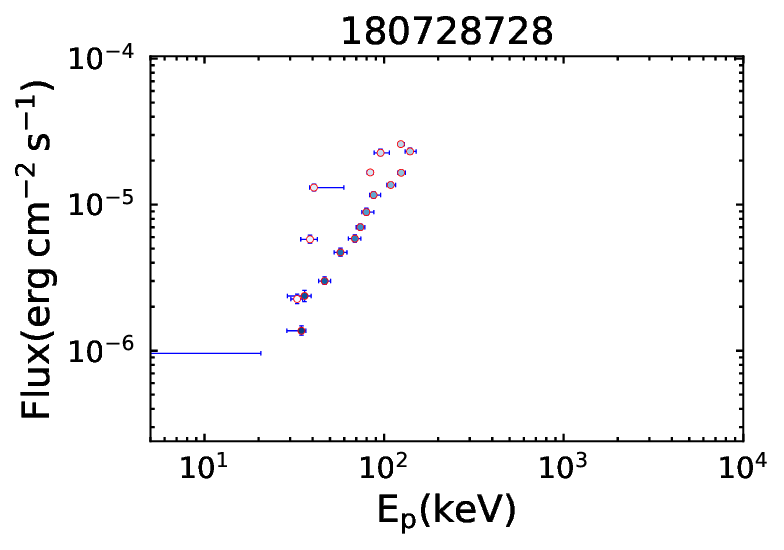}
     \includegraphics[width=\textwidth]{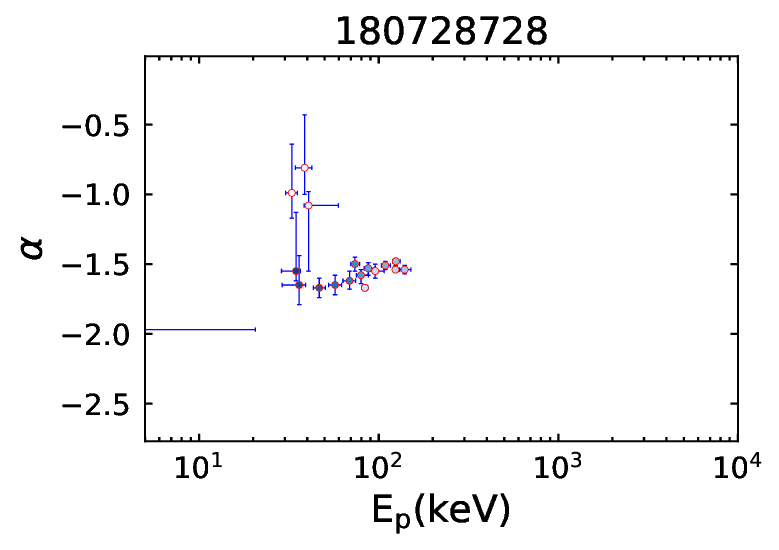}
     \end{minipage}
     \end{minipage}
      \begin{minipage}{0.49\textwidth}
          \begin{minipage}{0.6\textwidth}
	\includegraphics[width=\textwidth]{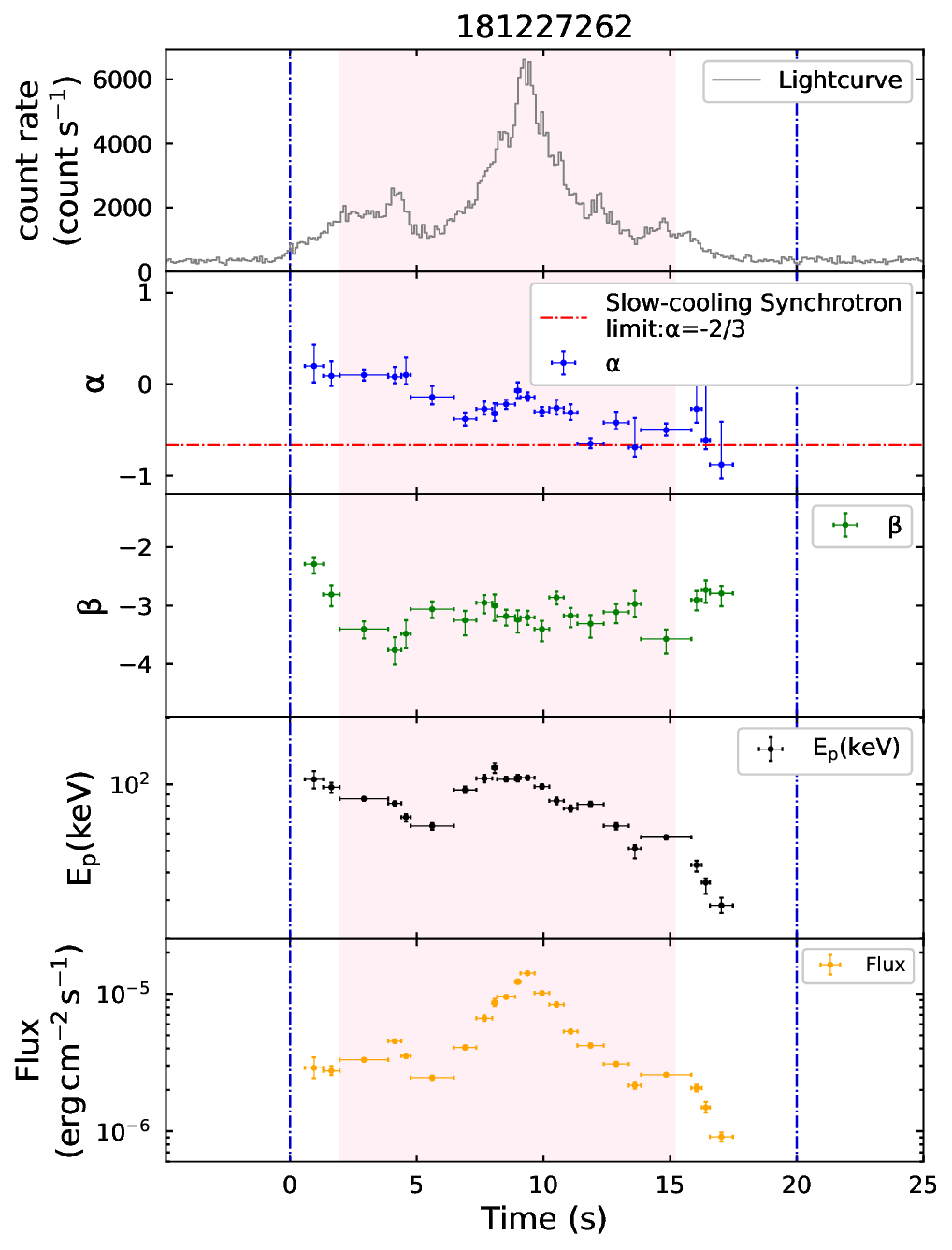}
         \end{minipage}
         \begin{minipage}{0.38\textwidth}
     \includegraphics[width=\textwidth]{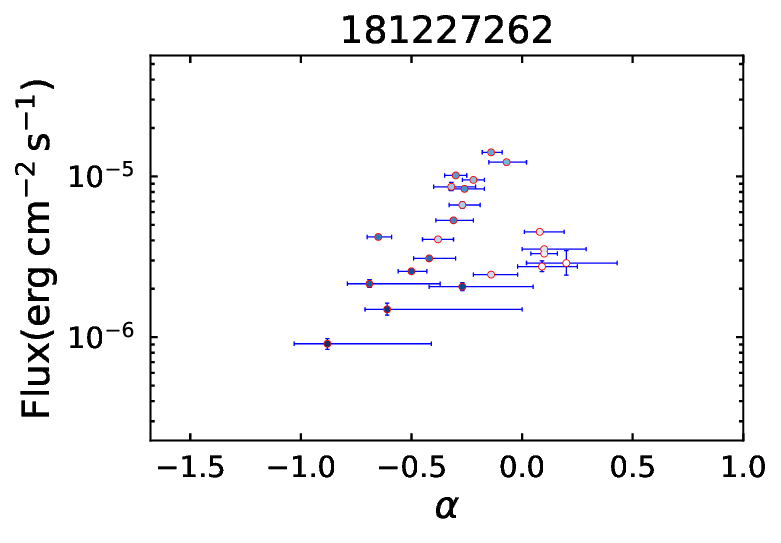}
     \includegraphics[width=\textwidth]{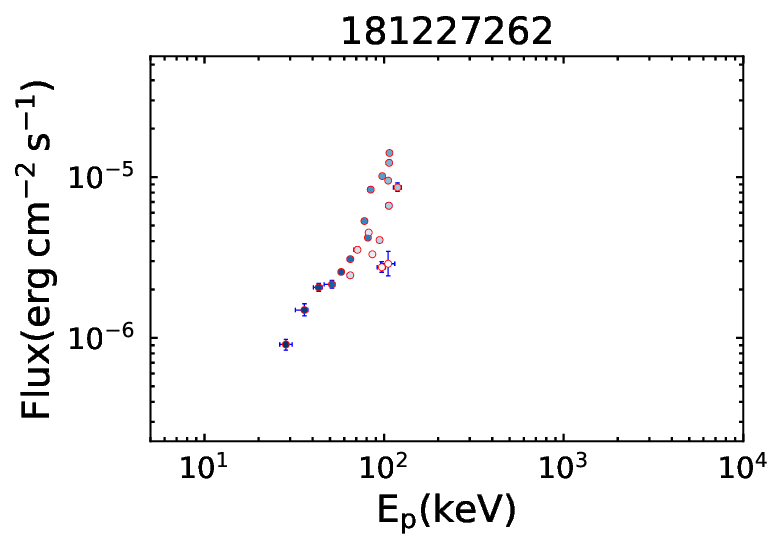}
     \includegraphics[width=\textwidth]{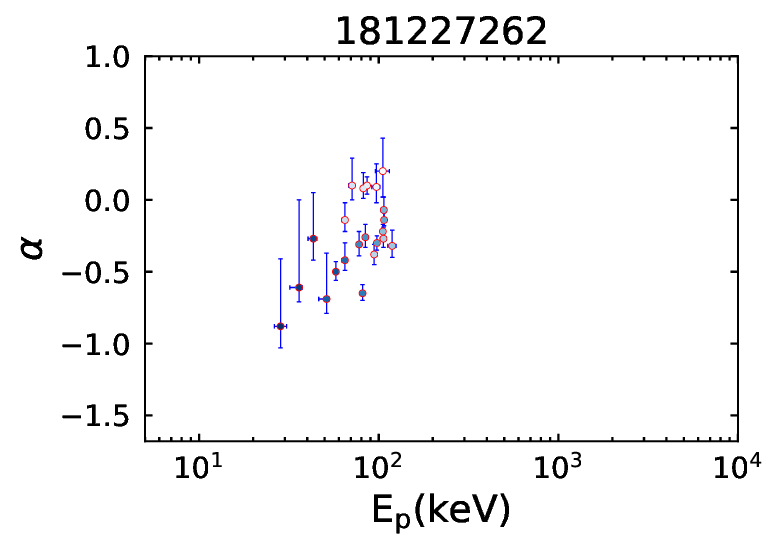}
     \end{minipage}
     \end{minipage} 
      \begin{minipage}{0.49\textwidth}
          \begin{minipage}{0.6\textwidth}
	\includegraphics[width=\textwidth]{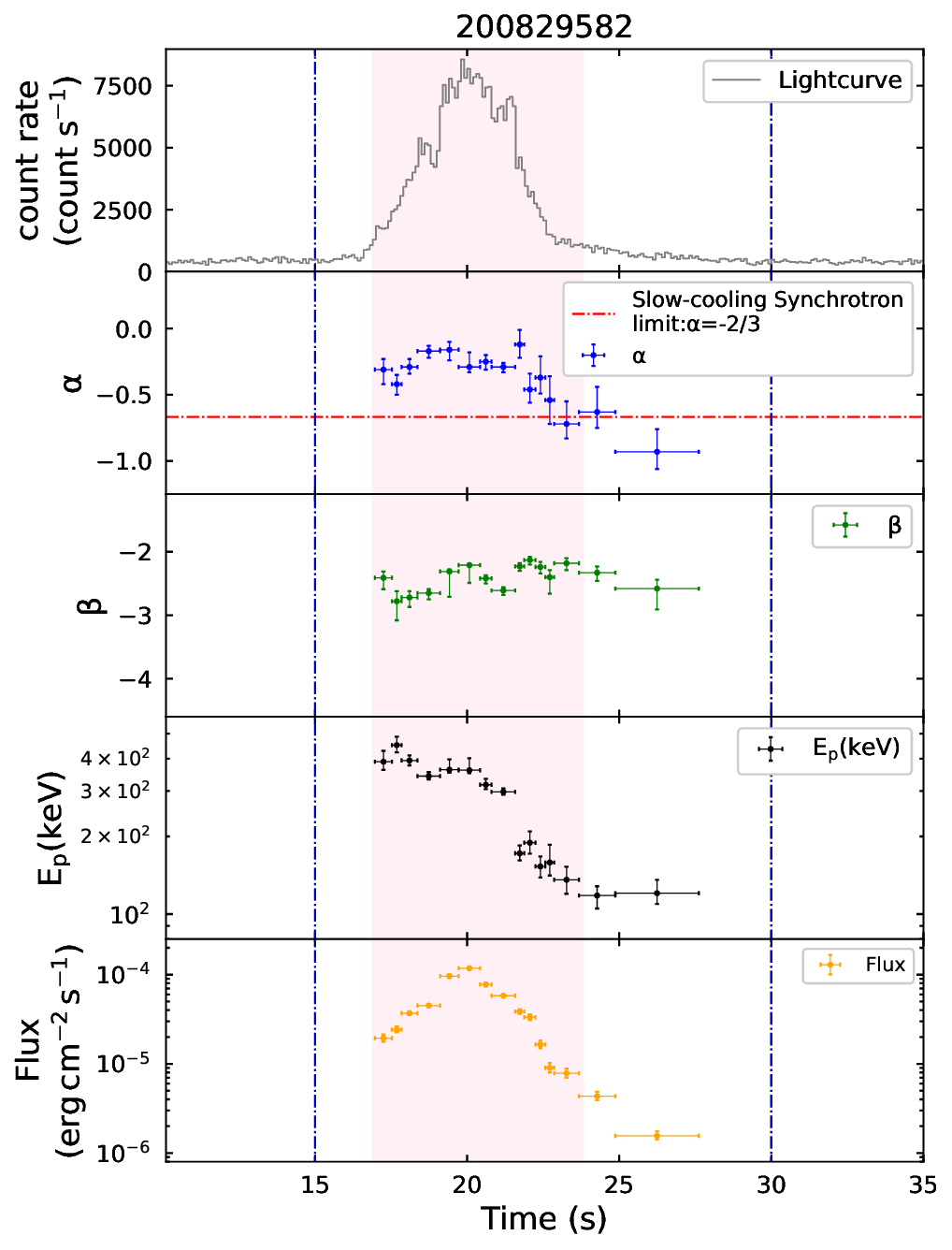}
         \end{minipage}
         \begin{minipage}{0.38\textwidth}
     \includegraphics[width=\textwidth]{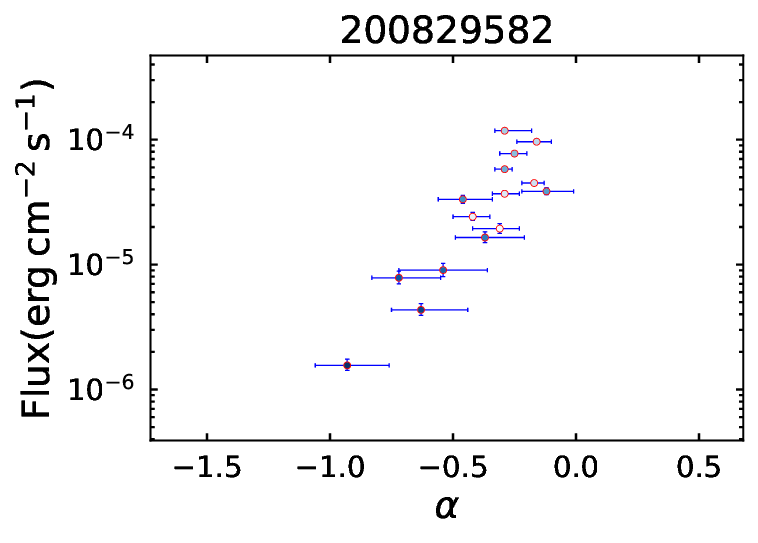}
     \includegraphics[width=\textwidth]{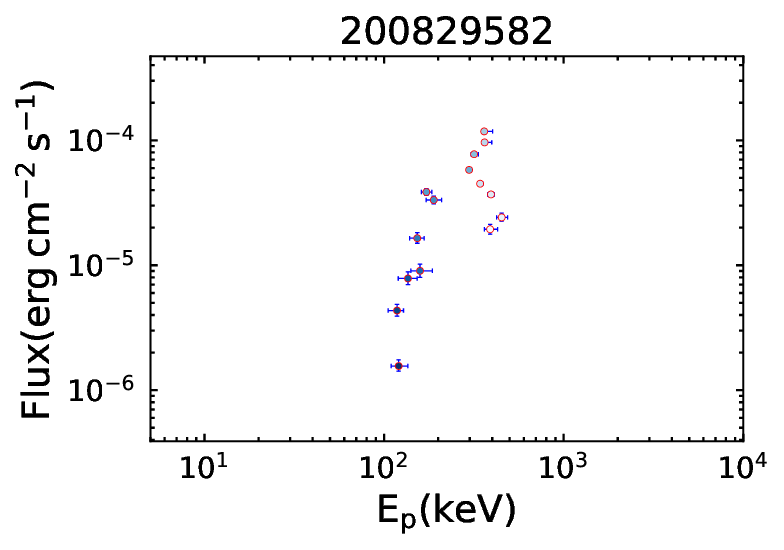}
     \includegraphics[width=\textwidth]{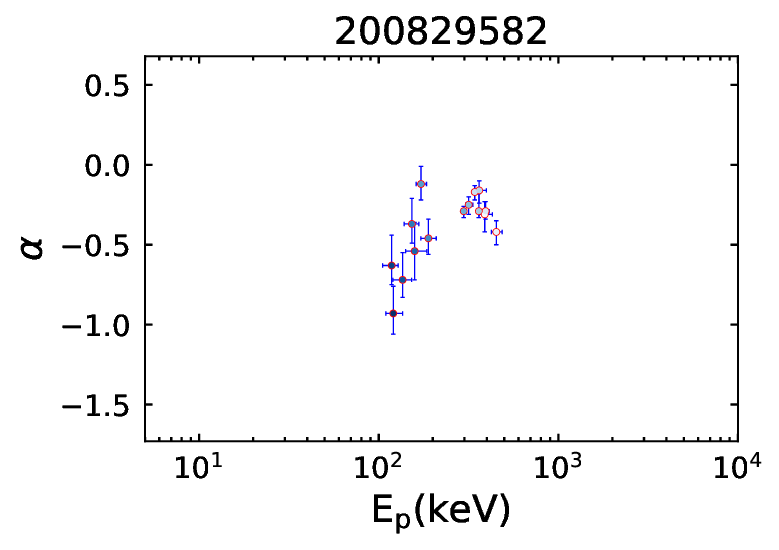}
     \end{minipage}
     \end{minipage}
      \begin{minipage}{0.49\textwidth}
          \begin{minipage}{0.6\textwidth}
	\includegraphics[width=\textwidth]{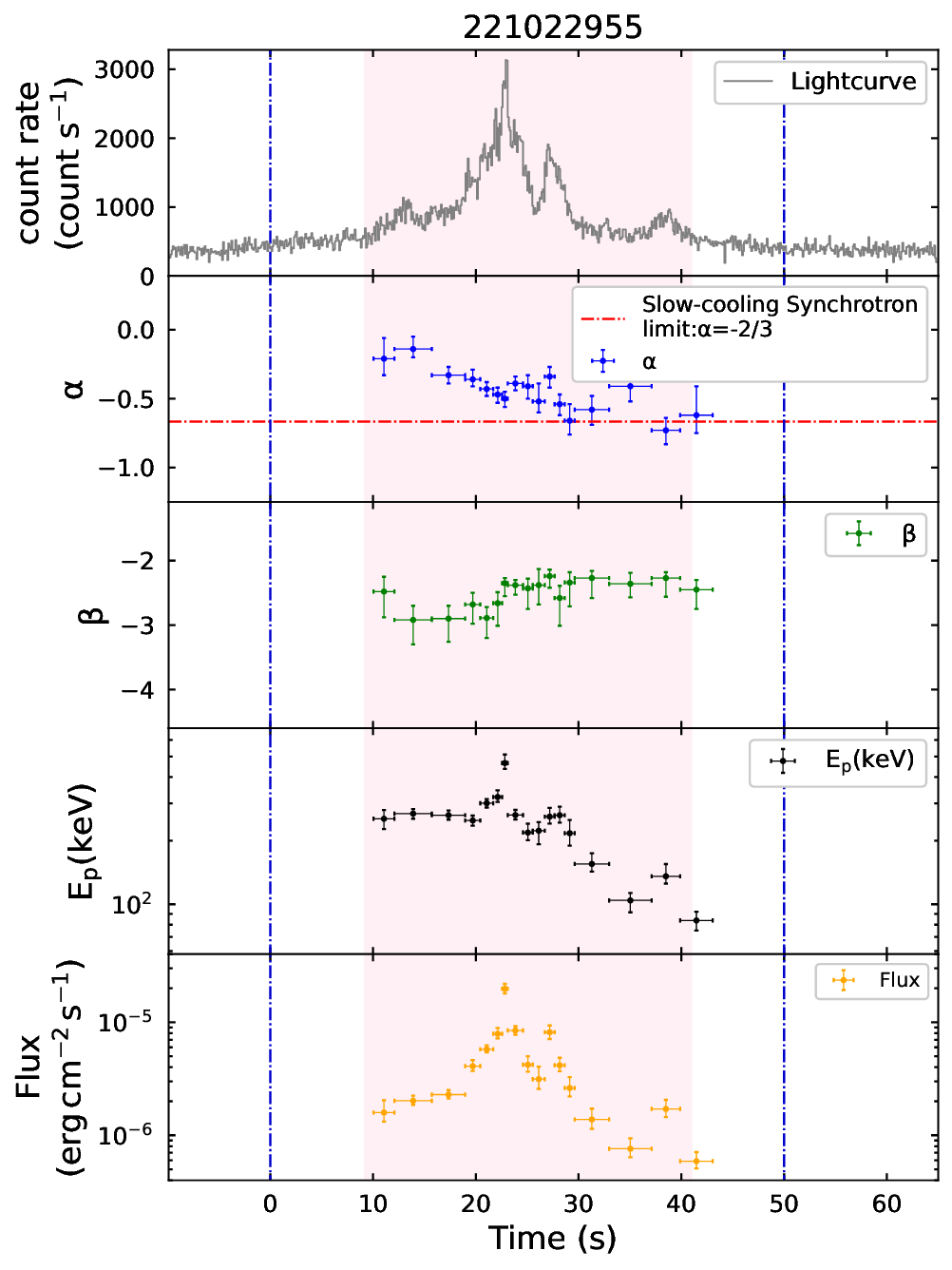}
         \end{minipage}
         \begin{minipage}{0.38\textwidth}
     \includegraphics[width=\textwidth]{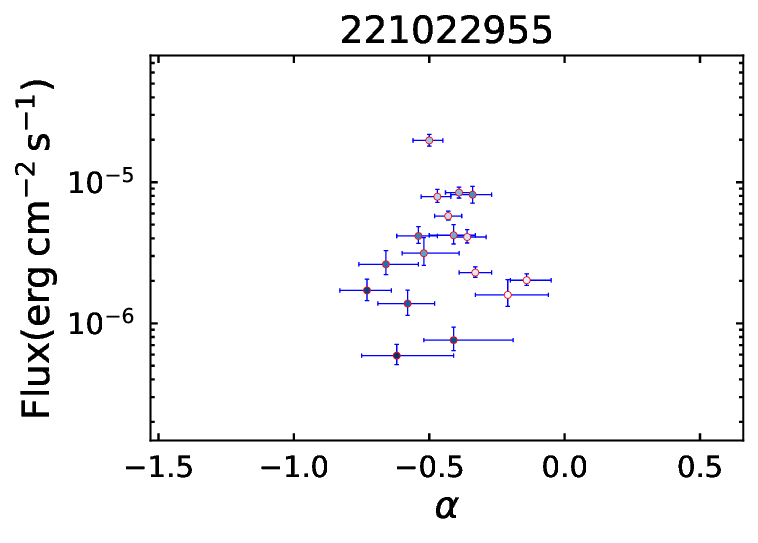}
     \includegraphics[width=\textwidth]{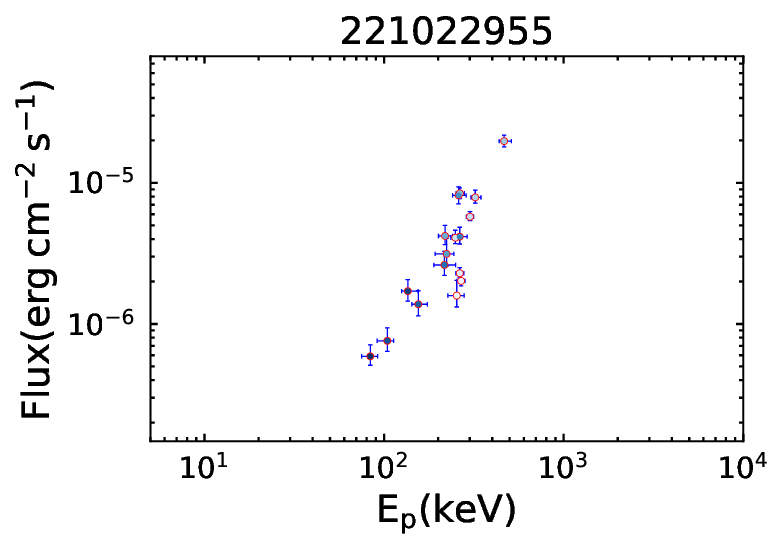}
     \includegraphics[width=\textwidth]{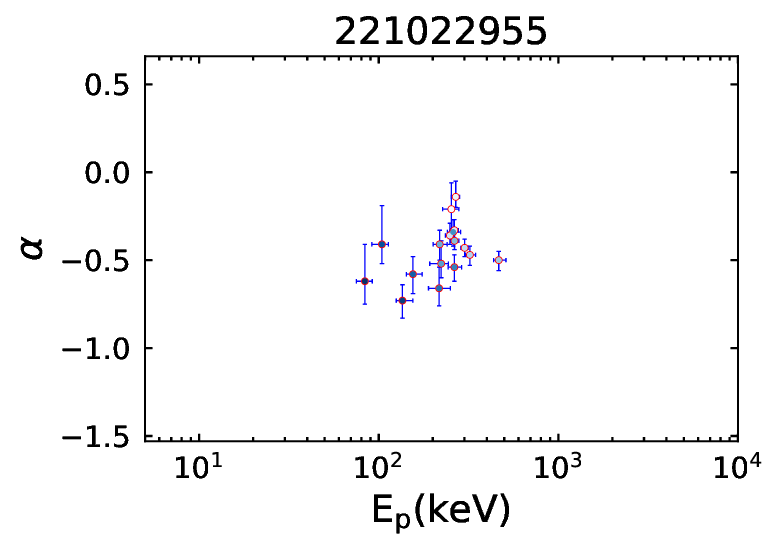}
     \end{minipage}
     \end{minipage}
      \begin{minipage}{0.49\textwidth}
          \begin{minipage}{0.6\textwidth}
	\includegraphics[width=\textwidth]{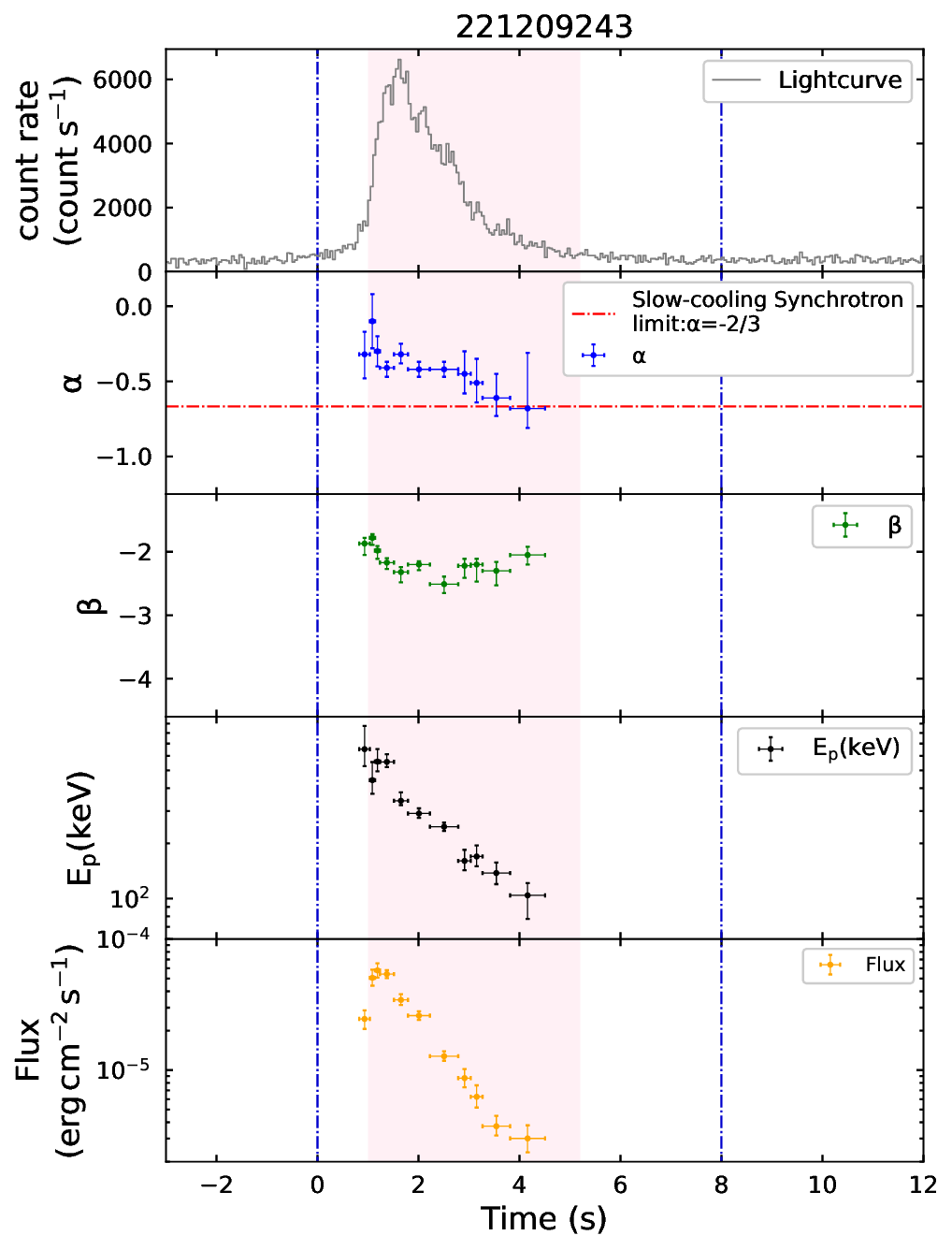}
         \end{minipage}
         \begin{minipage}{0.38\textwidth}
     \includegraphics[width=\textwidth]{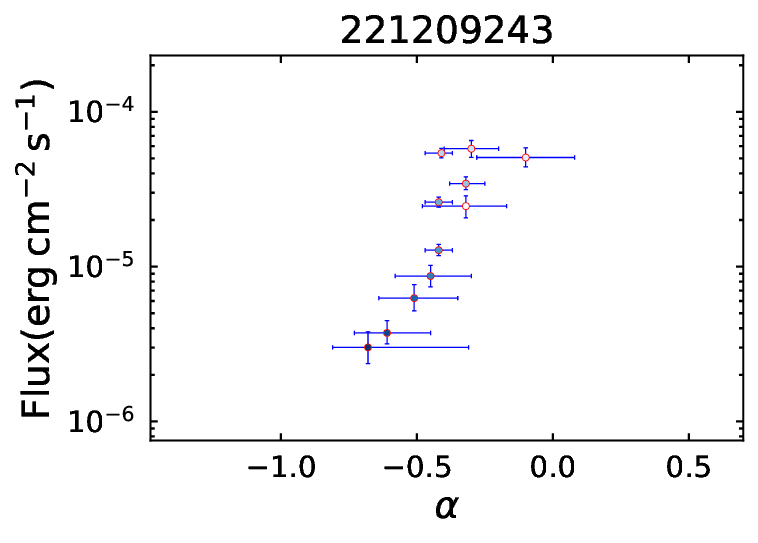}
     \includegraphics[width=\textwidth]{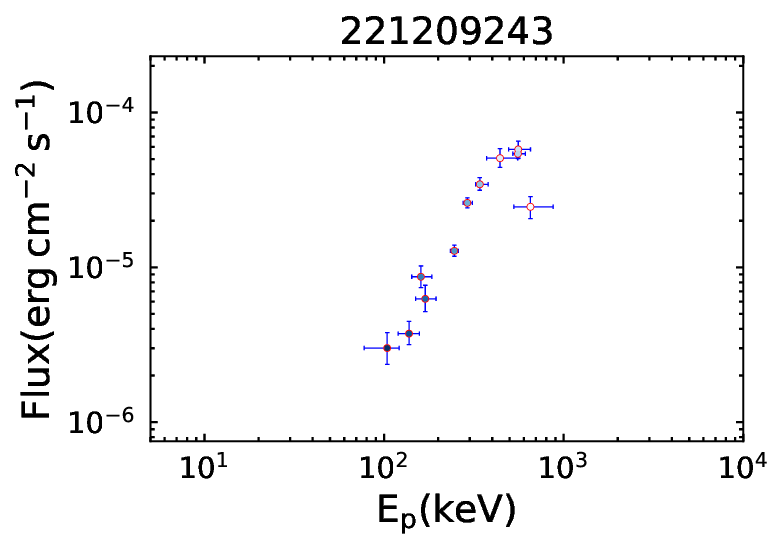}
     \includegraphics[width=\textwidth]{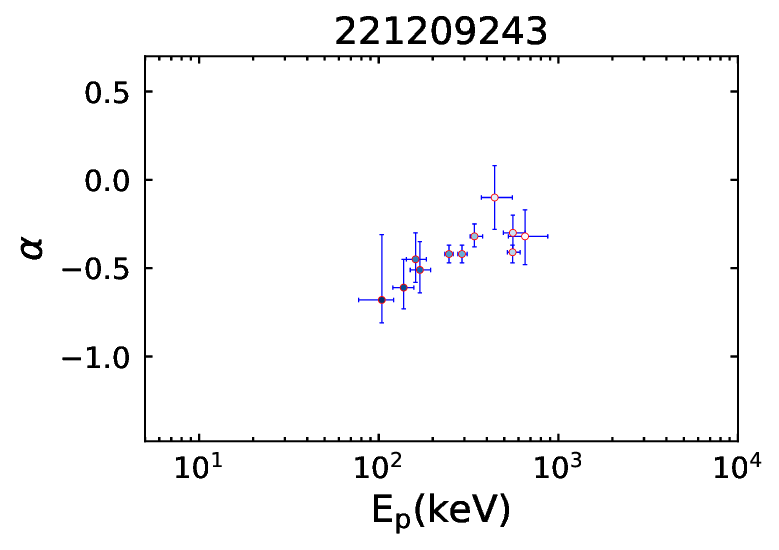}
     \end{minipage}
     \end{minipage} 
	\caption{Left: Illustration of parameter evolution, the pink shaded region representing the GBM $T_{90}$ period. The region contained by the blue dotted line is the selected source interval. Right: Relationships between $F-\alpha$, $F-E_{\rm p}$, and $\alpha-E_{\rm p}$ depicted. The color scale ranging from light blue (start) to deep blue (end) indicates temporal evolution.}
	\label{fig3:spec evolution}
\end{figure*}

\begin{figure*}
	\centering
	\includegraphics[width=0.48\textwidth]{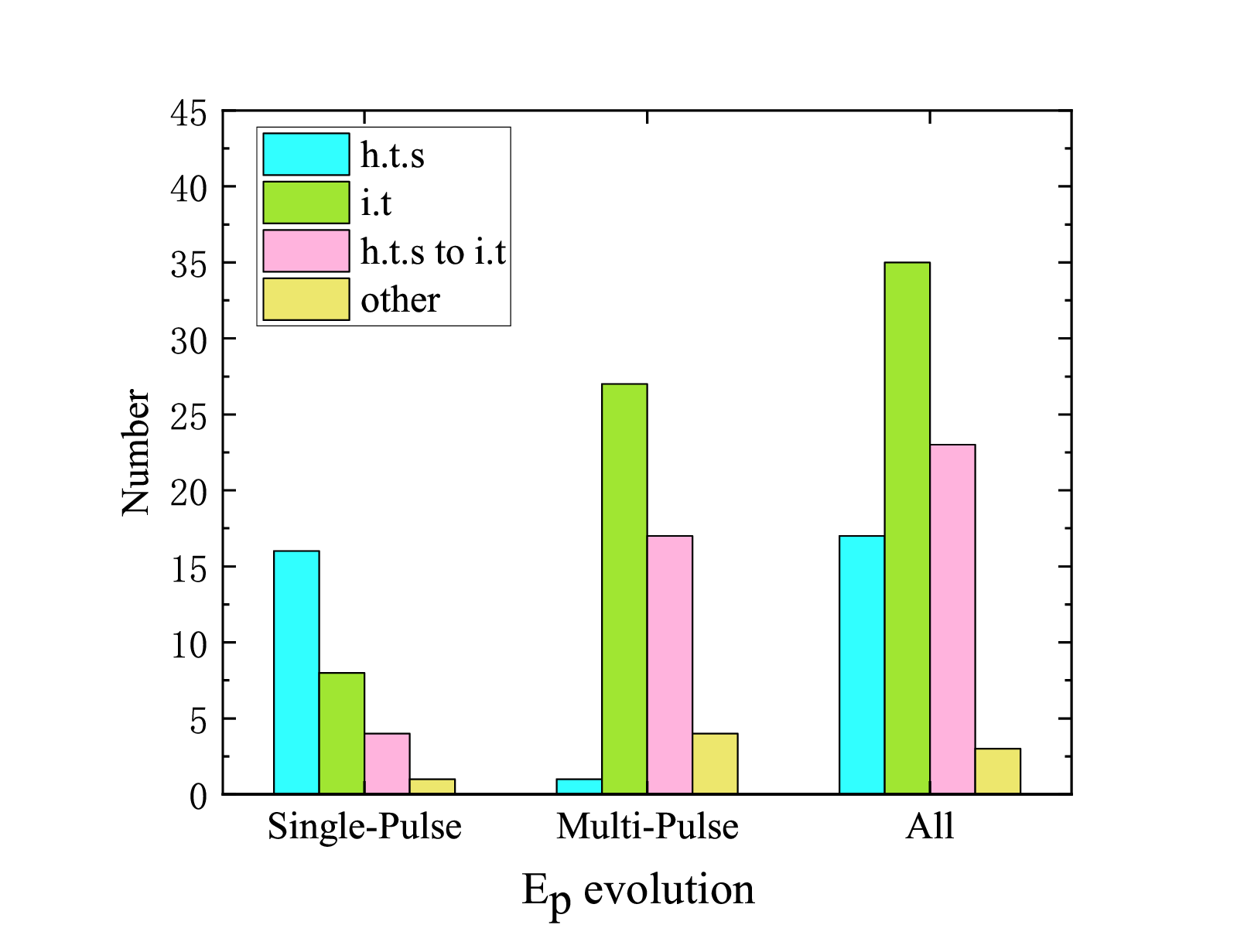}
     \includegraphics[width=0.48\textwidth]{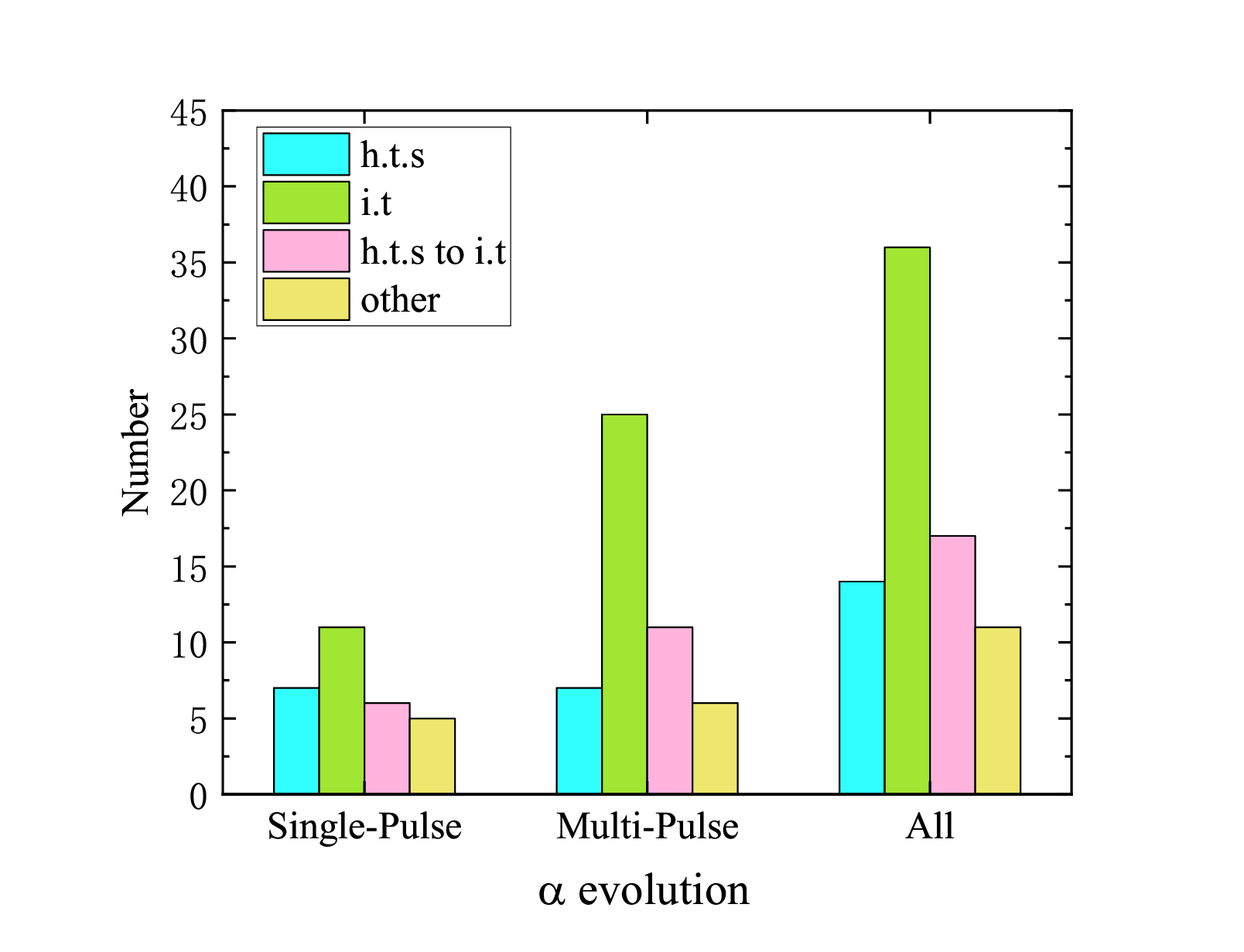}
	\caption{Histogram of parameter evolution, left panel displays the temporal evolution of $E_{\rm p}$, while right panel illustrates the evolution of $\alpha$. The first group shows single-pulse GRBs, the second group corresponds to multiple-pulse GRBs, and the third group encompasses all types. The abbreviation `` h.t.s" denotes a hard-to-soft pattern, ``i.t" signifies an intensity-tracking pattern, ``h.t.s to i.t" indicates a hard-to-soft followed by intensity-tracking pattern, and ``other" encompasses other evolution patterns.}
	\label{fig4:par evolution}
\end{figure*}

The evolution of spectral parameters, such as $\alpha$, $\beta$, $E_{\rm p}$ and $\nu F_{\nu}$ flux for 6 GRBs as representative examples, is illustrated in Figure \ref{fig3:spec evolution}. Based on the observed trends along with the lightcurve, we categorized the evolution patterns of $\alpha$ and $E_{\rm p}$ for all samples. Specifically, the patterns were classified into different categories: hard-to-soft (h.t.s), intensity-tracking (i.t), hard-to-soft followed by intensity-tracking (h.t.s to i.t), and other patterns. In this classification, ``hard" and ``soft" denote larger and smaller values of $\alpha$ and $E_{\rm p}$ parameters, respectively. The detailed results can be found in columns 3 and 4 of Table \ref{tab3:evolution and correlations}. The overall statistics for the evolution patterns are summarized in Figure \ref{fig4:par evolution} and Table \ref{tab4:par evolution}. \par

Regarding the evolution of $E_{\rm p}$ among our total sample of 78 long GRBs, the intensity-tracking pattern was exhibited by 35 bursts (45\%), the hard-to-soft followed by intensity-tracking pattern was displayed by 23 bursts (29\%), the hard-to-soft pattern was shown by 17 bursts (22\%), while three bursts (4\%) exhibited other evolution patterns. The dominant evolution patterns for $E_{\rm p}$ were the hard-to-soft pattern and the intensity-tracking pattern. Interestingly, in some GRBs both of these patterns coexisted, resulting in the hard-to-soft followed by intensity-tracking pattern. \par
 
In terms of the evolution of $\alpha$, the observed patterns were characterized by a high degree of complexity and unpredictability. The majority of our sample showed tracking patterns, hard-to-soft patterns, and hard-to-soft followed by intensity-tracking patterns. The evolution of $\alpha$ was roughly assessed, with 36 bursts (46\%) displaying a rough intensity-tracking pattern, 14 bursts (18\%) showing the hard-to-soft evolution pattern, 17 bursts (22\%) exhibiting the hard-to-soft followed by intensity-tracking pattern, and 11 GRBs (14\%) in the sample demonstrating other evolution patterns. Specifically, we observed several cases with different evolution patterns, such as the soft-to-hard followed by intensity-tracking (e.g., GRB 140213A), the intensity-tracking followed by soft-to-hard (e.g., GRB 200125B), and one flat evolution pattern (GRB 110920). \par

It is worth noting that the underlying physics processes governing single-pulse and multi-pulse bursts may differ significantly, thus it is crucial to investigate the spectral evolution patterns separately for these bursts. In our sample data set, there are 29 single-pulse GRBs. For $E_{\rm p}$ evolutionary behavior within this subset, three main patterns emerged 16 bursts (55\%) exhibited the hard-to-soft evolution, 8 bursts (28\%) showed the intensity-tracking evolution, and 4 bursts (14\%) displayed the hard-to-soft followed by intensity-tracking evolution pattern. One burst (GRB 110721A) demonstrated a hard-to-soft evolution followed by soft-to-hard pattern. In regards to the evolution of $\alpha$ in single-pulse GRBs, 11 bursts (38\%) showed an intensity-tracking pattern, 6 bursts (21\%) displayed the hard-to-soft followed by intensity-tracking pattern, and 7 bursts (24\%) demonstrated the hard-to-soft trend.\par
 
Among the 49 multi-pulse GRBs, 27 bursts (55\%) exhibited an intensity-tracking pattern for the evolution of $E_{\rm p}$, while 17 bursts (35\%) demonstrated the hard-to-soft followed by intensity-tracking pattern. One burst (GRB 130704) exhibited the hard-to-soft pattern, and 4 bursts (8\%) displayed other evolution patterns. For the evolution of $\alpha$, 25 bursts (51\%) showed the intensity-tracking pattern. 11 bursts (23\%) showed the hard-to-soft followed by intensity-tracking pattern, with the hard-to-soft pattern usually appearing in the first pulse. 7 bursts (14\%) showed hard-to-soft pattern. These evolution patterns observed for both $\alpha$ and $E_{\rm p}$ align with previous studies (\citealt{Yu+etal+2019, Duan+etal+2020, Li+etal+2021}). \cite{Lu+etal+2012} suggested that all subsequent pulses exhibiting the intensity-tracking behavior were attributed to superimposed hard-to-soft pulses; however, it is noteworthy that the intensity-tracking pattern was also identified within the first pulse itself, indicating that this phenomenon represents a genuine characteristic of $E_{\rm p}$ evolution. 

\onecolumn
\bc
\topcaption{Evolution of Parameters  $E_{\rm p}$ and $\alpha$ and Parameter Relations
\label{tab3:evolution and correlations}}
\setlength{\tabcolsep}{10pt}
\small
\tablefirsthead{\hline\noalign{\smallskip} 
GRB          & $a>-2/3$     & $E_{\rm p}$     & $\alpha $      & $\alpha - E_{\rm p}$    &$ F-E_{\rm p}$    & $F-\alpha$          & pulse \\
 &          &                (Evolution)  & (Evolution)             & type(r)        & type(r)            & type(r)                  & (N) \\
\hline\noalign{\smallskip}}
\tablehead{
\multicolumn{8}{l}{{Continued from previous page}} \\ 
\hline\noalign{\smallskip}
GRB          & $a>-2/3$     & $E_{\rm p}$     & $\alpha $      & $\alpha - E_{\rm p}$    &$ F-E_{\rm p}$    & $F-\alpha$          & pulse \\
 &          &                (Evolution)  & (Evolution)             & type(r)        & type(r)            & type(r)                  & (N) \\
\hline\noalign{\smallskip}}
\tabletail{%
\multicolumn{8}{r}{{Continued on next page}} \\ \hline\noalign{\smallskip}}
\tablelasttail{\noalign{\smallskip}\hline}
\begin{supertabular}{cccccccc}
    081009140     & not all      & i.t          & i.t          & 2p(0.5)      & 2p(0.94)     & 1(0.45)      & 2 \\
    081125496     & yes          & h.t.s        & i.t          & 1(0.38)      & 2p(0.94)     & 2p(0.46)     & 1 \\
    081215784     & not all      & h.t.s to i.t & i.t          & 1(0.37)      & 2p(0.9)      & 2p(0.64)     & 3 \\
    081221681     & not all      & h.t.s to i.t & h.t.s to i.t & 2p(0.79)     & 2p(0.92)     & 1(0.71)      & 2 \\
    081224887     & not all      & h.t.s        & h.t.s        & 2p(0.94)     & 2p(1)        & 2p(0.94)     & 1 \\
    090719063     & not all      & h.t.s to i.t & h.t.s        & 2p(0.85)     & 2p(0.95)     & 2p(0.78)     & 1 \\
    090820027     & not all      & i.t          & h.t.s        & 2p(0.63)     & 2p(0.9)      & 1(0.73)      & 1 \\
    090902462     & not all      & i.t          & h.t.s to i.t & 3(0.01)      & 2p(0.87)     & 3(-0.01)     & 3 \\
    090926181     & not all      & i.t          & i.t          & 3(0.32)      & 2p(0.76)     & 3(0.28)      & 2 \\
    091127976     & not all      & h.t.s to i.t & i.t          & 1(0.53)      & 2p(0.39)     & 2p(0.45)     & 3 \\
    100324172    & not all      & h.t.s to i.t & h.t.s        & 2p(0.64)     & 2p(0.82)     & 1(0.57)      & 2 \\
    100707032    & not all      & h.t.s        & i.t          & 2p(0.95)     & 2p(0.97)     & 2p(0.99)     & 1 \\
    100719989    & not all      & i.t          & h.t.s to i.t & 1(0.28)      & 2p(0.8)      & 2p(0.48)     & 3 \\
    101123952    & not all      & h.t.s to i.t to h.t.s & i.t & 3(0.48)      & 2p(0.82)     & 2p(0.64)     & 5 \\
    101126198    & no           & i.t          & i.t to s.t.h & 2p(0.55)     & 2p(0.89)     & 1(0.54)      & 1 \\
    110301214    & not all      & h.t.s to i.t & h.t.s to i.t & 2p(0.64)     & 2p(0.75)     & 1(0.55)      & 2 \\
    110625881    & not all      & h.t.s to i.t & i.t          & 2p(0.53)     & 2p(0.79)     & 3(0.46)      & 3 \\
    110721200    & no           & h.t.s to s.t.h & i.t        & 1(0.02)      & 1(0.52)      & 1(0.67)      & 1 \\
    110920546    & yes          & h.t.s        & flat         & 2n(-0.7)     & 2p(0.98)     & 3(-0.62)     & 1 \\
    111220486    & no           & h.t.s to i.t & i.t          & 3(0.19)      & 2p(0.65)     & 3(0.37)      & 2 \\
    120119170    & no           & h.t.s to i.t & h.t.s to i.t & 2p(0.68)     & 2p(0.76)     & 2p(0.5)      & 1 \\
    120328268    & not all      & h.t.s to i.t & i.t          & 2p(0.64)     & 2p(0.85)     & 2p(0.75)     & 2 \\
    120711115    & no           & i.t          & h.t.s to i.t & 3(-0.15)     & 3(0.21)      & 2p(0.67)     & 2 \\
    120728434    & not all      & i.t          & $\cdots$     & 2n(-0.92)    & 2p(0.79)     & 2n(-0.61)    & 5 \\
    120919309    & not all      & i.t          & h.t.s to i.t & 2n(0.67)     & 2p(0.9)      & 2p(0.5)      & 1 \\
    130518580    & not all      & i.t          & h.t.s to i.t & 1(0.37)      & 2p(0.91)     & 3(0.46)      & 1 \\
    130606497    & not all      & i.t          & h.t.s to i.t & 3(-0.41)     & 2p(0.87)     & 3(-0.14)     & 4 \\
    130704560    & not all      & h.t.s        & i.t          & 2n(0.89)     & 1(0.73)      & 2p(0.85)     & 3 \\
    131014215    & not all      & i.t          & h.t.s        & 2n(0.44)     & 2p(0.75)     & 1(0.23)      & 2 \\
    140206275    & not all      & h.t.s to i.t & h.t.s to i.t & 2n(0.83)     & 1(0.88)      & 2p(0.89)     & 2 \\
    140213807    & not all      & i.t          & s.t.h to i.t & 3(0.09)      & 2p(0.85)     & 1(0.21)      & 2 \\
    140329295    & not all      & h.t.s to i.t & i.t          & 3(0.2)       & 2p(0.78)     & 1(0.44)      & 2 \\
    141028455    & not all      & h.t.s        & i.t to s.t.h & 1(0.33)      & 1(0.58)      & 2p(0.64)     & 1 \\
    150127589    & not all      & i.t to h.t.s & h.t.s        & 2p(0.91)     & 2p(0.57)     & 1(0.57)      & 2 \\
    150201574    & not all      & i.t          & i.t          & 2p(0.7)      & 2p(0.89)     & 2p(0.9)      & 2 \\
    150330828    & not all      & i.t          & h.t.s to i.t & 2p(0.66)     & 2p(0.65)     & 1(0.49)      & 5 \\
    150403913    & no           & h.t.s to i.t & i.t          & 2p(0.46)     & 1(0.60)      & 2p(0.93)     & 2 \\
    150902733    & not all      & h.t.s to i.t & i.t          & 1(0.30)      & 2p(0.63)     & 2p(0.66)     & 1 \\
    160113398    & yes          & h.t.s        & i.t          & 1(0.51)      & 1(0.81)      & 2p(0.84)     & 1 \\
    160530667    & not all      & i.t          & i.t          & 2p(0.7)      & 2p(0.89)     & 2p(0.8)      & 1 \\
    160905471    & not all      & i.t          & i.t          & 2p(0.64)     & 2p(0.92)     & 2p(0.71)     & 2 \\
    160910722    & not all      & h.t.s        & i.t          & 2p(0.79)     & 1(0.49)      & 2p(0.70)     & 1 \\
    170405777    & not all      & i.t          & i.t to h.t.s & 3(0.29)      & 2p(0.66)     & 3(0.48)      & 3 \\
    170522657    & yes          & h.t.s to i.t & h.t.s        & 1(0.43)      & 1(0.26)      & 2n(-0.54)    & 2 \\
    170808936    & not all      & i.t          & i.t          & 1(0.81)      & 2p(0.88)     & 2p(0.82)     & 3 \\
    170826819    & not all      & i.t          & h.t.s        & 2p(0.43)     & 2p(0.84)     & 2n(0.55)     & 3 \\
    171210493    & not all      & h.t.s        & s.t.h to i.t & 1(-0.25)     & 2p(1)        & 3(-0.25)     & 1 \\
    171227000    & not all      & i.t          & h.t.s to i.t & 2p(0.63)     & 2(0.93)      & 2p(0.53)     & 3 \\
    180113418    & not all      & h.t.s to i.t & flat to s.t.h & 2n(-0.19)    & 1(0.69)      & 2n(-0.34)    & 2 \\
    180305393    & yes          & h.t.s        & h.t.s to i.t & 2n(0.71)     & 2p(0.77)     & 3(0.38)      & 1 \\
    180720598    & no           & h.t.s to i.t & i.t          & 1(0.5)       & 2p(0.8)      & 2p(0.72)     & 5 \\
    180728728    & no           & i.t          & h.t.s        & 3(0.16)      & 2p(0.91)     & 3(0.24)      & 1 \\
    181227262    & not all      & h.t.s to i.t & h.t.s to i.t & 2p(0.48)     & 2p(0.83)     & 1(0.30)      & 2 \\
    190114873    & not all      & i.t          & flat to i.t  & 2p(0.87)     & 2p0.95)      & 1(0.89)      & 3 \\
    190530430    & not all      & i.t          & i.t          & 3(0.4)       & 2p(0.81)     & 2p(0.77)     & 3 \\
    190720613    & no           & h.t.s to i.t & i.t          & 1(0.65)      & 2p(0.83)     & 2p(0.74)     & 3 \\
    190727846    & no           & i.t          & i.t          & 1(0.42)      & 2p(0.70)     & 3(0.5)       & 4 \\
    190731943    & not all      & h.t.s to i.t & h.t.s to i.t & 1(0.42)      & 2p(0.71)     & 2p(0.88)     & 1 \\
    200101861    & not all      & i.t          & i.t          & 1(0.4)       & 2p(0.68)     & 2p(0.79)     & 2 \\
    200125864    & not all      & i.t          & i.t to s.t.h & 2p(0.5)      & 2p(0.88)     & 2p(0.61)     & 5 \\
    200313071    & not all      & h.t.s        & i.t to s.t.h & 1(0.32)      & 2p(0.70)     & 1(0.75)      & 1 \\
    200412381    & not all      & i.t          & i.t          & 1(0.32)      & 2p(0.81)     & 2p(0.70)     & 2 \\
    200826923    & not all      & i.t          & i.t          & 2p(0.51)     & 1(0.61)      & 3(0.58)      & 1 \\
    200829582    & not all      & h.t.s        & i.t          & 1(0.55)      & 1(0.62)      & 2p(0.86)     & 1 \\
    201016019    & not all      & h.t.s        & i.t          & 1(0.68)      & 1(0.78)      & 2p(0.93)     & 1 \\
    201216963    & no           & i.t          & i.t          & 3(0.31)      & 2p(0.73)     & 3(0.31)      & 2 \\
    210406949    & no           & h.t.s to i.t & i.t          & 2p(0.89)     & 1(0.61)      & 2p(0.58)     & 3 \\
    210610827    & yes          & i.t          & h.t.s        & 2p(0.54)     & 2p(0.94)     & 3(0.42)      & 3 \\
    210714331    & not all      & h.t.s        & i.t          & 3(0.35)      & 2p(0.66)     & 2p(0.79)     & 1 \\
    210801581    & no           & i.t          & i.t          & 2p(0.60)     & 2p(0.71)     & 2p(0.52)     & 2 \\
    211019250    & not all      & h.t.s        & h.t.s        & 2p0.95)      & 2p(0.98)     & 2p(0.97)     & 1 \\
    220304228    & yes          & h.t.s        & h.t.s        & 2p(0.9)      & 2p(0.76)     & 2p(0.64)     & 1 \\
    220426285    & yes          & h.t.s to i.t & h.t.s to i.t & 2p(0.86)     & 1(0.75)      & 1(0.63)      & 2 \\
    220527387    & not all      & i.t          & i.t          & 3(0.33)      & 2p(0.59)     & 2p(0.82)     & 4 \\
    220910242    & not all      & h.t.s to i.t & i.t          & 2p(0.49)     & 2p(0.87)     & 2p(0.60)     & 3 \\
    221022955    & not all      & i.t          & h.t.s        & 2p(0.44)     & 2p(0.74)     & 3(0.16)      & 2 \\
    221023862    & no           &  i.t         & h.t.s to i.t & 3(0.23)      & 2p(0.92)     & 2p(0.420     & 1 \\
    221209243    & not all      & h.t.s        & h.t.s        & 2p(0.89)     & 2p(0.85)     & 2p(0.89)     & 1 \\
\end{supertabular}
\tablecomments{0.86\textwidth}{Spectral proportions of $\alpha$ across synchrotron death lines (column 2), $E_{\rm p}$ evolution (column 3), $\alpha$ evolution (column 4), types of parameter relations and Spearman rank coefficients (columns 5 to 7) were analyzed.}
\ec
\twocolumn

\begin{table*}
    \bc
\begin{threeparttable}
  \small
  \setlength{\tabcolsep}{20pt}
  \caption{Statistical Results of $E_{\rm p}$ and $\alpha$ Evolution}
  \label{tab4:par evolution}
    \begin{tabular}{cccccc}
    \hline\noalign{\smallskip}
     GRB & Sample & h.t.s & i.t   & h.t.s to i.t & other\tnote{a} \\
      & (N\tnote{b} ) & N(PCT\tnote{c} ) & N(PCT) & N(PCT) & N(PCT) \\
    \hline\noalign{\smallskip}
    $E_{\rm p}$ Evolution &       &       &       &       &  \\
    \hline\noalign{\smallskip}
    Single-Pulse & 29    & 16(55\%) & 8(28\%) & 4(14\%) & 1(3\%) \\
    Multi-Pulse & 49    & 1(2\%) & 27(55\%) & 17(35\%) & 4(8\%) \\
    All   & 78    & 17(22\%) & 35(45\%) & 23(29\%) & 3(4\%) \\
    \hline\noalign{\smallskip}
    $\alpha$ Evolution &       &       &       &       &  \\
    \hline\noalign{\smallskip}
    Single-Pulse & 29    & 7(24\%) & 11(38\%) & 6(21\%) & 5(17\%) \\
    Multi-Pulse & 49    & 7(14\%) & 25(51\%) & 11(23\%) & 6(12\%) \\
    All   & 78    & 14(18\%) & 36(46\%) & 17(22\%) & 11(14\%) \\
    \noalign{\smallskip}\hline
    \end{tabular}
\begin{tablenotes}
    \footnotesize
    \item[a] Hard-to-soft (h.t.s); intensity-tracking (i.t); hard-to-soft followed by intensity-tracking (h.t.s to i.t); other patterns (other).
    \item[b] Number of samples.
    \item[c] Percentage.
\end{tablenotes}
\end{threeparttable}
   \ec
\end{table*}

\subsection{Global Parameter Relations}

\begin{figure*}
	\centering
	\includegraphics[width=0.48\textwidth]{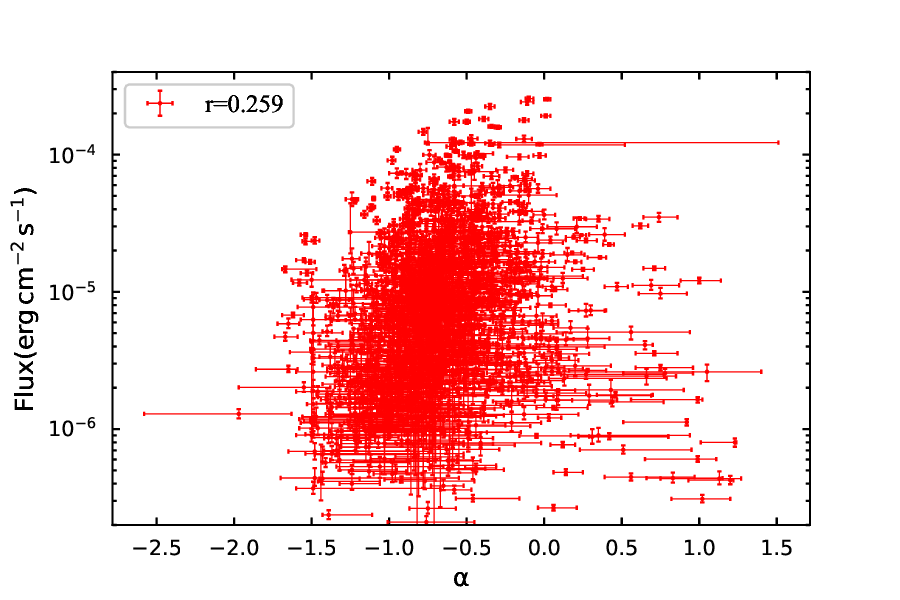}
        \includegraphics[width=0.48\textwidth]{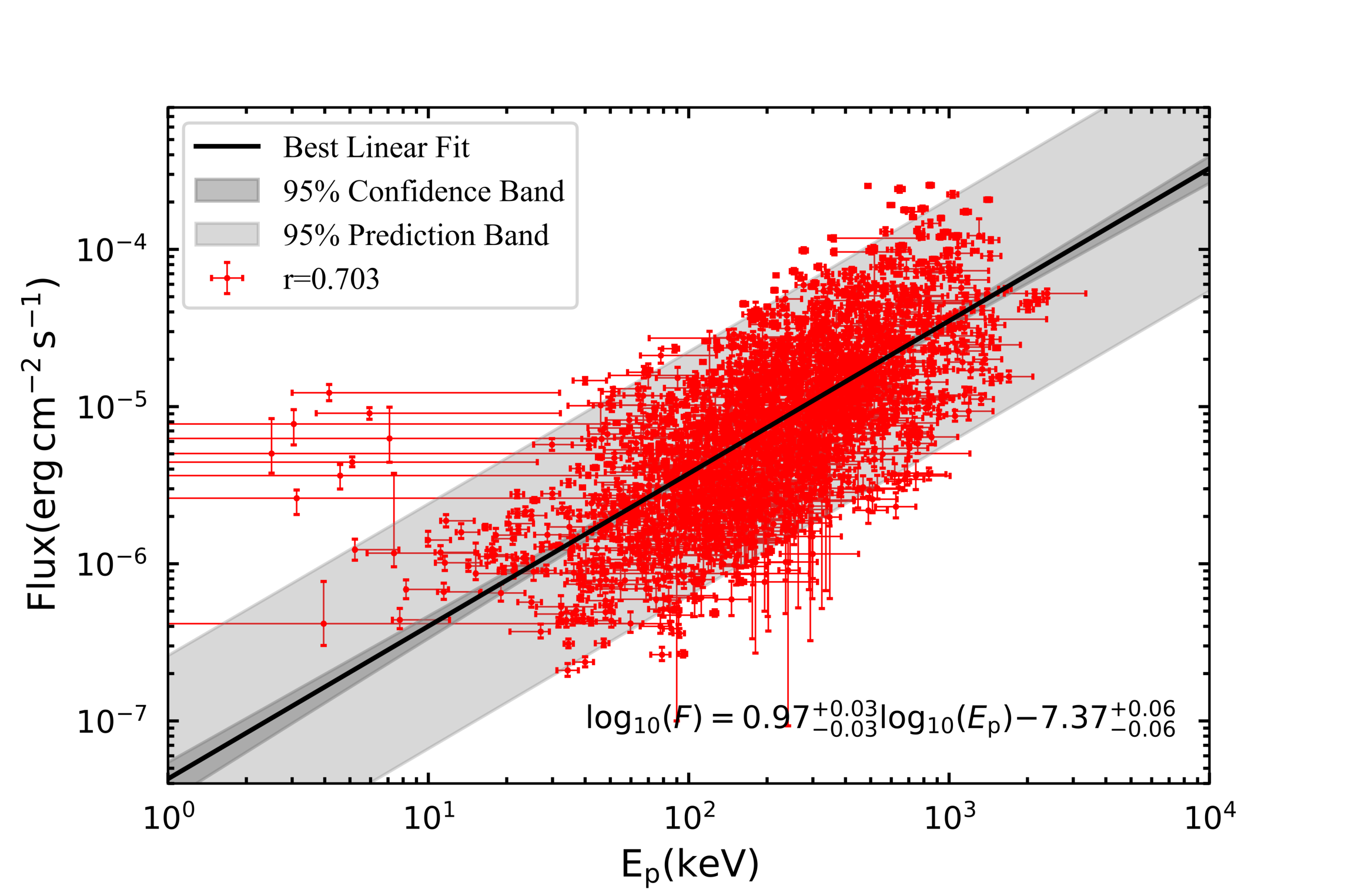}
        \includegraphics[width=0.48\textwidth]{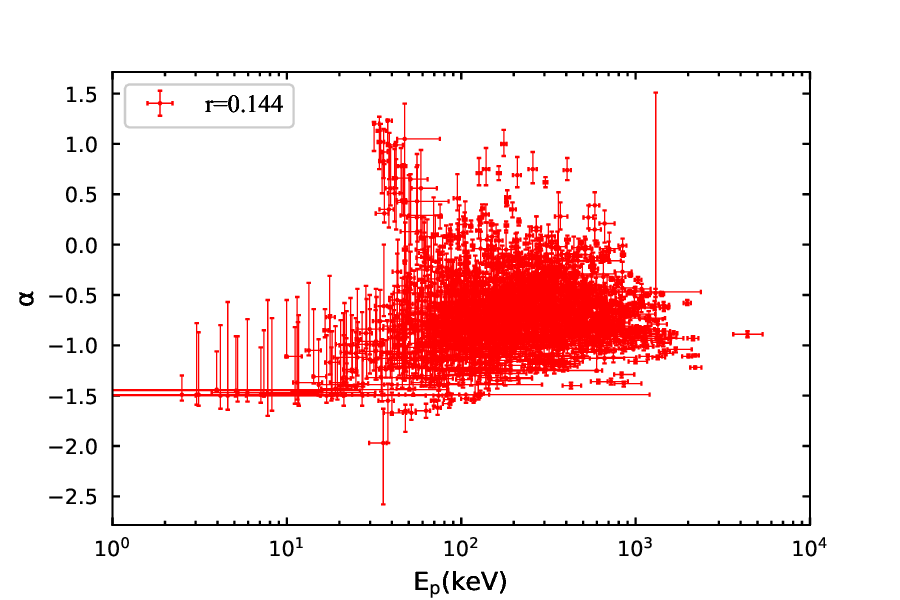}
        \includegraphics[width=0.48\textwidth]{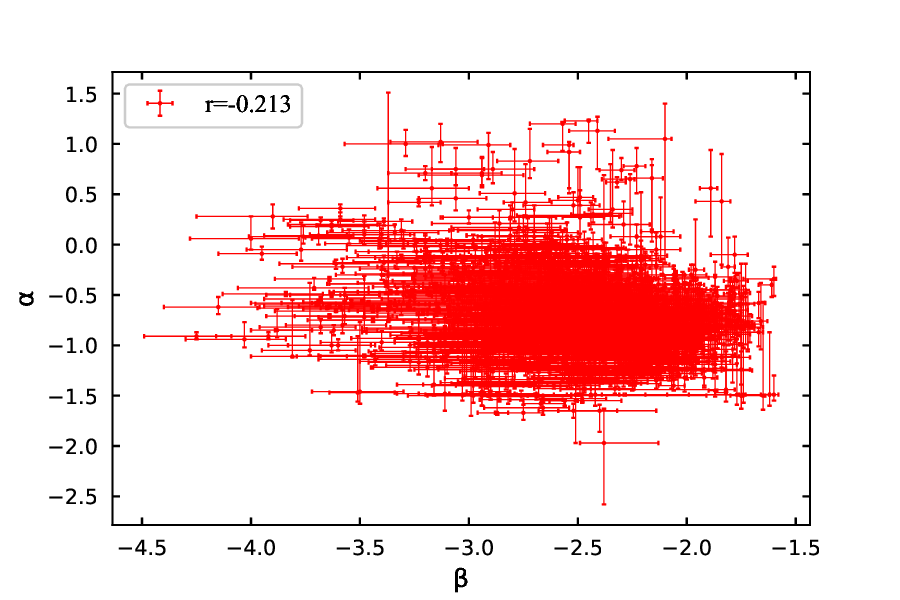}
	\caption{Global relationships between spectra parameters with statistical significance ${S \geq 20}$, the $F-\alpha$  relation (top left panel), the $F-E_{\rm p}$ relation (top right panel), the $\alpha-E_{\rm p}$ relation (bottom left panel), and the $\alpha-\beta$ relation (bottom right panel).}
	\label{fig5:global relation}
\end{figure*}

The correlations among parameters play a crucial role in understanding the characteristics of prompt radiation from GRBs. Previous studies have primarily focused on investigating relationships between various parameters. An early discovery was the Golenetskii correlation, which established a strong correlation between flux ($F$) and peak energy ($E_{\rm p}$) (\citealt{Golenetskii+etal+1983}). In this study, we further examined the global correlations among selected sample parameters. The overall relationships between four groups of parameters are depicted in Figure \ref{fig5:global relation}: the $F-\alpha$ relation (top left panel), the $F-E_{\rm p}$ relation (top right panel), the $\alpha-E_{\rm p}$ relation (bottom left panel), and the $\alpha-\beta$ relation (bottom right panel). We quantified these parameter correlations using the Spearman correlation coefficient ($r$), where $r > 0.7$ indicates a strong correlation, and $r < 0.4$ indicates a weak correlation. \par

\begin{figure}
	\centering
	\includegraphics[width=0.48\textwidth]{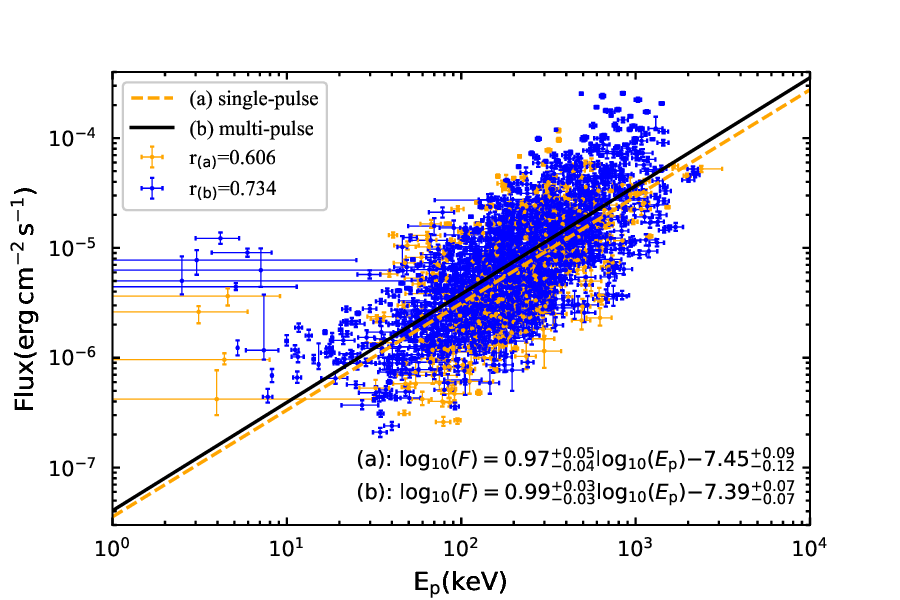}
	\caption{The global $F-E_{\rm p}$ relationship for single-pulse and multi-pulse GRBs, orange and blue data points represent single-pulse and multi-pulse GRBs, respectively. The fitted lines are represented by dashed and solid lines correspondingly. Spearman's correlation rank coefficients for single-pulse ($r_{\rm (a)}$) and multi-pulse ($r_{\rm (b)}$) GRBs were calculated.}
	\label{fig6:F_Ep_classify_relation}
\end{figure}

Our results indicate a strong positive correlation for the $F-E_{\rm p}$ relation in the log-log plot, with a correlation coefficient of $r = 0.702\,(p<10^{-4})$, thus confirming the validity of the Golenetskii correlation. The fitted   $F-E_{\rm p}$ relationship is given as  
\begin{equation}\label{eq2}
{\log_{10}F = 0.97^{+0.03}_{-0.03}\log_{10}(E_{\rm p})-7.37^{+0.06}_{-0.06}},
\end{equation}

\noindent where the units of $E_{\rm p}$ and $F$ are keV and $\rm erg\,cm^{-2}\,s^{-1}$, respectively. Additionally, we also separately analyzed the $F-E_{\rm p}$ relationship for single-pulse and multi-pulse GRBs, shown in Figure \ref{fig6:F_Ep_classify_relation}, resulting in two distinct relationships:
\begin{equation}\label{eq3}
 \begin{array}{ll}
 {\log_{10}F = 0.97^{+0.05}_{-0.04}\log_{10}(E_{\rm p})-7.45^{+0.09}_{-0.12}} \text{\ (single),} \\ \\
 {\log_{10}F = 0.99^{+0.03}_{-0.03}\log_{10}(E_{\rm p})-7.37^{+0.07}_{-0.07}} \text{\ (multi).} 
 \end{array}
\end{equation}

\noindent Their Spearman rank correlation coefficients are $r = 0.606\,(p<10^{-4})$ and $r = 0.734\,(p<10^{-4})$, respectively. It can be observed that both single-pulse and multi-pulse GRBs exhibit power-law exponents $\sim 1$ for their respective $F-E_{\rm p}$ relationships without any apparent difference. The absolute values of the correlation coefficients for the other three relations are below 0.4, indicating weak correlations.

\subsection{Individual Parameter Relations}
\begin{figure}
	\centering
	\includegraphics[width=0.45\textwidth]{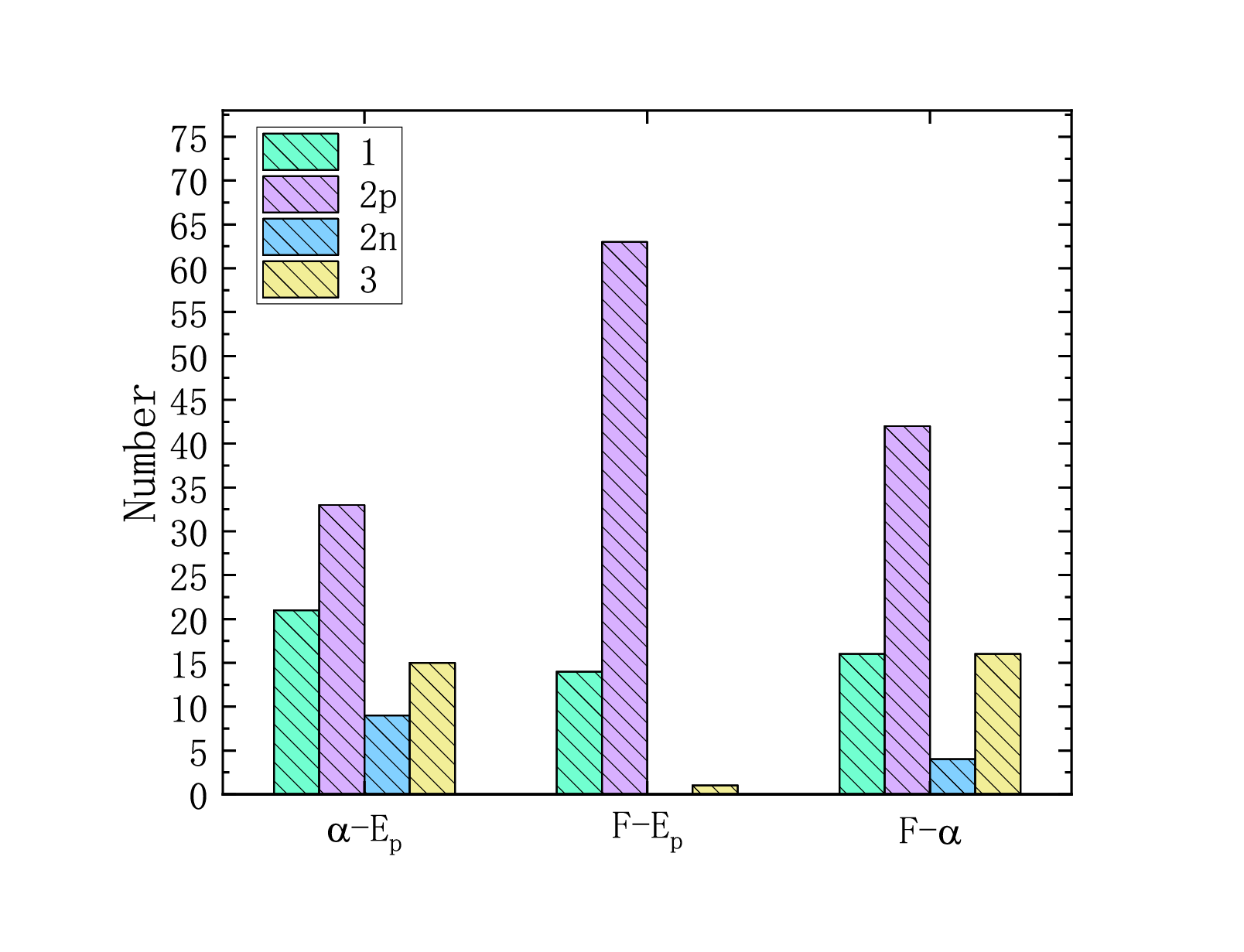}
        \includegraphics[width=0.42\textwidth]{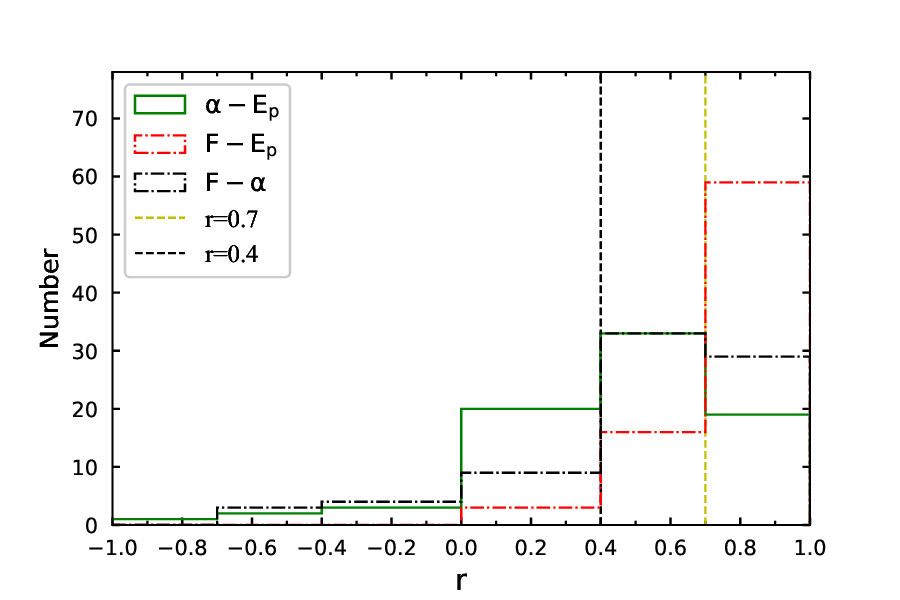}
	\caption{Top panel: histograms illustrating the parameter relations types, specifically $\alpha-E_{\rm p}$,  $F-E_{\rm p}$ , and $F-\alpha$ relations for groups 1, 2, and 3, respectively. Bottom panel: distribution histograms of Spearman's rank coefficient $r$ for the parameter relations.}
	\label{fig7:statistical relation}
\end{figure}

We conducted a comprehensive analysis of parameter relationships for each GRB, and some examples are presented in Figure \ref{fig3:spec evolution}. Following the categorization scheme proposed by \cite{Yu+etal+2019}, we classified the individual relationships into three categories: (i) non-monotonic relations (type 1), which include both positive and negative power-law segments; (ii) monotonic relations that can be described by a single power law, including monotonic positive correlations (type 2p) and monotonic negative correlations (type 2n); and (iii) relationships exhibiting no apparent trend (type 3). We quantified these parameter relationships using Spearman's rank coefficients. The specific relation types and corresponding Spearman coefficients for each burst are summarized in columns 5 to 7 of Table \ref{tab3:evolution and correlations}. The distribution of relation types and correlation coefficients for individual bursts is illustrated in Figure \ref{fig7:statistical relation}.

\subsubsection{\texorpdfstring{$F-E_{\rm p}$}{F-Ep} relation}

From the results presented in Figure \ref{fig3:spec evolution}, Table \ref{tab3:evolution and correlations}, and Figure \ref{fig7:statistical relation}, it can be observed that the relationship between flux ($F$) and peak energy ($E_{\rm p}$) in GRBs predominantly exhibits either a monotonic correlation (type 2) or a non-monotonic correlation (type 1) when plotted on a logarithmic scale. Among the individual $F-E_{\rm p}$ relationships, 63 bursts (81\%) are categorized as monotonically positively correlated (type 2p), while 14 bursts (18\%) demonstrate a non-monotonic trend (type 1), additionally, one burst does not exhibit apparent trend (type 3). Within the sample set, strong correlations ($0.7<r<1$) are observed in 59 cases (76\%), moderate correlations ($0.4<r<0.7$) are found in 16 instances (20\%), while 3 bursts (4\%) indicate weak correlations ($0<r<0.4$).

\subsubsection{\texorpdfstring{$F-\alpha$}{F-α} relation}

Regarding the $F-\alpha$ relation, it is observed that 42 bursts (53\%) exhibit a monotonically increasing trend (type 2p), while 4 bursts (5\%) display a monotonically decreasing trend (type 2n). Additionally, 16 bursts (21\%) demonstrate a non-monotonic trend (type 1), and another 16 bursts (21\%) do not exhibit a clear trend pattern (type 3). Analyzing Spearman’s rank coefficient reveals that out of these bursts, 29 bursts (37\%) show a strong correlation, whereas, for 33 bursts (42\%), there is a moderate correlation. Furthermore, in the case of 9 bursts (12\%) show correlations,  7 bursts (9\%) display negative associations. Notably, seven bursts are found to have a negative correlation, out of which three bursts demonstrate a moderately negative correlation ($-0.7 < r < -0.4$) and four bursts show a weakly negative correlation ($-0.4 < r < 0$).

\subsubsection{\texorpdfstring{$\alpha-E_{\rm p}$}{α-Ep} relation}

Regarding the $\alpha-E_{\rm p}$ relation, it is observed that 33 bursts (42\%) exhibit a monotonically positive correlation when plotted on a linear-log scale (type 2p), while 9 bursts (12\%) display a monotonically negative correlation (type 2n), Additionally, 21 bursts (27\%) show non-monotonic relationships (type 1), and for 15 bursts (19\%), do no significant trend is demonstrated (type 3). Analyzing the statistical results of the correlation coefficients, it is revealed that 19 bursts (24\%) exhibit strong correlations, 33 bursts (42\%) show moderate correlations, 20 bursts (26\%) display weak correlations, 6 bursts (8\%) demonstrate negative correlations ($-1<r<0$). Notably, GRB 120728B shows a strong negative correlation ($r = -0.92$). It can be observed that the $\alpha-E_{\rm p}$ relation does not have a clearly dominant type of relationship, and most of the samples exhibit relatively weak correlations.
\par

\section{Summary and Discussion}
\label{sect:discussion}
In this work, we performed a detailed time-resolved spectral analysis of 78 bright long GRBs detected by Fermi/GBM. Our selected sample yielded 1490 time-resolved spectra that satisfy the statistical significance ${S \geq 20}$. The Band function was employed to fit all time-resolved spectra. Firstly, We statistically analyzed the parameters ($\alpha$, $\beta$, $E_{\rm p}$), the derived parameter ($F$) of all time-resolved spectra as well as the maximum (hardest) of the low-energy index. Secondly, we investigated the evolution patterns of both the peak energy $E_{\rm p}$ and the low-energy index $\alpha$. Finally, we examined correlations among parameter relations $F-E_{\rm p}$, $F-\alpha$, and $\alpha-E_{\rm p}$. \par

The fitted results of the time-resolved spectral parameters are as follows: $\alpha=-0.72 \pm 0.32$, $\beta=-2.42 \pm 0.39$, $\mathrm{log_{10}}{(E_{\rm p}\rm{/keV})}=\mathrm{log_{10}{(221.69)} \pm 0.41}$, $\mathrm{log_{10}}{(F\rm{/(erg\,cm^{-2}\,s^{-1}})}=\mathrm{log_{10}{(7.49e-6)} \pm 0.59}$. The value of $\alpha$ in our sample is slightly harder (-0.72) than the typical value of the previous works ($-0.8$) but remains within the synchrotron limit. The hardest low-energy index $\alpha_{\rm max}$ in each burst has 86\% surpasses the synchrotron limit. \par

As for the distribution of spectral evolution patterns, there exists a slight disparity between multi-pulse GRBs and single-pulse GRBs. For multi-pulse GRBs, one can conclude that the intensity-tracking pattern is more common than the hard-to-soft pattern for the evolution of both the peak energy $E_{\rm p}$ and the low-energy index $\alpha$. For single-pulse GRBs, the evolution pattern of peak energy $E_{\rm p}$ was more common for the hard-to-soft than the intensity-tracking pattern. It should be pointed out that the universally existed intensity-tracking pattern of $E_{\rm p}$ in multi-pulse GRBs might be a natural aspect of the positive relationship of $F-E_{\rm p}$ showed before.\par

The parameter relationship between the global sample and the individual samples was also investigated, revealing a robust positive correlation between $E_{\rm p}$ and $F$ in the entire data set. In the individual samples, a significant majority (81\%) of the $F-E_{\rm p}$ relations exhibit a monotonic positive correlation, indicating an intrinsic association between flux and peak energy. Additionally, approximately half (53\%) of the $F-\alpha$ relations exhibit a monotonic positive correlation. The parameter correlations have been investigated in some previous works, the $F-E_{\rm p}$ relationship has been found to be strongly correlated both in the time-integrated spectra of large GRB samples (e.g., \citealt{Golenetskii+etal+1983, Borgonovo+etal+2001}) and the time-resolved ones of an individual GRB (e.g., \citealt{Lu+etal+2012, Yu+etal+2019, Li+etal+2021}). The $F-\alpha$ relationship was found to be positively correlated in some GRB time-resolved spectra (\citealt{Yu+etal+2019, Ryde+etal+2019, Li+etal+2021}). Our results are generally consistent with the study of 103 pulses from 38 multi-pulse GRBs by \cite{Li+etal+2021}. For the $F-E_{\rm p}$ relationship, they found that 74 pulses (71\%) exhibited monotonic positive correlations, and for the $F-\alpha$ relationship, they found that 69 pulses (67\%) exhibited monotonic positive correlations.\par

The Band component of most observed gamma-ray burst spectra is widely believed to originate from synchrotron radiation. There are at least two models proposed for interpreting the prompt emission of GRBs: the internal shocks (IS) model, as described by (\citealt{Paczynskietal+1986, Goodmanetal+1986, Shemietal+1990, Reesetal+1992, Meszarosetal+1993, Reesetal+1994}) and the Internal-collision-induced Magnetic Reconnection and Turbulence (ICMART) model, as described by (\citealt{Zhangbin+etal+2011}). Both models are capable of producing synchrotron radiation. In the internal shock model, the peak energy $E_{\mathrm{p}} \propto \gamma_{e}^{2} L^{1 / 2} R^{-1}(1+z)^{-1}$ (\citealt{Zhang+etal+2002}), where $L$ is the ``wind" luminosity of the ejecta, $\gamma_{e}$ is the electron Lorentz factor in the emission region, $R$ is the emission radius, and $z$ is the redshift of the gamma-ray burst. Despite predicting a positive correlation between peak energy ($E_{\rm p}$) and luminosity (flux), the power law index of the theoretical model does not align with the observed $F-E_{\rm p}$ relations. The positive correlation of the $F-\alpha$ relation can be attributed to subphotospheric heating within the flow with varying entropy as explained by \cite{Ryde+etal+2019}. According to their findings, during the peak phase of the peak of the light curve, a high entropy causes the photosphere to approach its saturation radius, leading to a narrow spectrum with intense emission. Conversely, during the pulse decay phase when entropy decreases, weaker emission and broader spectrum are expected as the photosphere moves away from its saturation radius. 

As mentioned above, the hard-to-soft pattern of $E_{\rm p}$ evolution is abundantly present in single-pulse GRBs and the first pulse of multi-pulse GRB, and the intensity-tracking pattern of $E_{\rm p}$ evolution is present in a fraction of single-pulse GRBs and a large number of multi-pulse GRBs. Both two evolution patterns have already been predicted by different emission models. The IS model can give a tracking behavior, since $E_{\rm p} \propto L^{1 / 2}$ (\citealt{Zhang+etal+2002}). A hard-to-soft evolution pattern of $E_{\rm p}$ evolution is predicted by the ICMART model (\citealt{Zhangbin+etal+2011}). On the other hand, the photosphere model can also reproduce an $E_{\rm p}$-tracking pattern (\citealt{Deng+etal+2014}). In partial multi-pulse GRBs, the hard-to-soft pattern and the intensity-tracking pattern of $E_{\rm p}$ evolution coexist in different pulses of a same GRB, and there is a transition between these two evolution patterns. As for the physical interpretation, two different possibilities could be drawn from the coexistence and transition of two evolution patterns in a same GRB. One is that two different emission mechanisms separately produce two $E_{\rm p}$ evolution patterns and transfer from one to the other. Another point of view is that only one complicated emission model (which considers finer physical details) can also produce the coexistence and transition of two evolution patterns. \cite{Uhm+etal+2018} demonstrated that a synchrotron model within a bulk-accelerated emission region can successfully reproduce both a hard-to-soft evolution and an intensity-tracking pattern of $E_{\rm p}$. Furthermore, \cite{Gao+etal+2021} demonstrated through numerical simulations that the synchrotron model can achieve a tracking pattern of $E_{\rm p}$ in two cases, one is that the cooling process of electrons is dominated by adiabatic cooling or synchrotron self-Compton (SSC, see \citealt{Derishev+etal+2001, Geng+etal+2018})+adiabatic cooling at the same time, the other is that the emitting region is accelerated and dominated by SSC cooling. Otherwise, a hard-to-soft pattern is normally expected. Additionally, the photospheric emission model from a structured jet can reproduce a hard-to-soft pattern and an intensity-tracking pattern of $E_{\rm p}$ evolution (\citealt{Meng+etal+2019}). In addition, \cite{Shao+etal+2022} showed that the ICMART model can produce the hard-to-soft pattern and the intensity-tracking pattern of $E_{\rm p}$ evolution. For one ICMART event, its intrinsic evolution of $E_{\rm p}$ is a hard-to-soft pattern. Considering one observed single-pulse could be formed by overlapping many sub-pulses produced by multiple ICMART events, the resultant evolution of $E_{\rm p}$ can exhibit an intensity-tracking pattern. In multi-pulse GRB produced by multiple ICMART events, the two evolution patterns of $E_{\rm p}$ can coexist with a variety of pattern transitions. This scenario can also reproduce most of the $E_{\rm p}$ evolution patterns found in this paper. Noting the majority of hard-to-soft pattern in sing-pulse GRBs, the ICMART seems to be a very competitive model. In summary, the diversity of spectral evolution patterns indicates that there may be more than one radiation mechanism occurring in the gamma-ray burst radiation process, including photospheric radiation and synchrotron radiation. However, it may also involve only one radiation mechanism, but more complicated physical details need to be considered.

\normalem
\begin{acknowledgements}

This work was performed under the auspices of the Science and Technology Foundation of Guizhou Province (grant No. QianKeHeJiChu ZK[2021]027), Major Science and Technology Program of Xinjiang Uygur Autonomous Region through No. 2022A03013-1, National Key Research and Development Program of China (No. 2022YFC2205202), and the National Natural Science Foundation of China grants 12288102, 12041304 and 11847102. We acknowledge the use of the Fermi archive’s public data.

\end{acknowledgements}

\bibliographystyle{raa}
\bibliography{bibtex}

\end{document}